\documentclass[manuscript,screen,nonacm]{acmart}
\usepackage{array} 
\usepackage{placeins} 
\newcommand{\appnote}[1]{\par\vspace{6pt}{\small\leftskip=0pt\rightskip=0pt\parfillskip=0pt plus 1fil\textit{Note.} #1\par}}

\AtBeginDocument{%
  }

\setcopyright{none}

\begin{document}

\title{Personalized to Persuade: The Effects of Contextualization and Warmth on Trust and Reliance in Conversational AI}
\renewcommand{\shorttitle}{Personalized to Persuade}


\author{Mert Yazan}
\orcid{0009-0004-3866-597X}
\affiliation{
 \institution{Amsterdam University of Applied Sciences}
  \city{Amsterdam}
 \country{Netherlands}
}
\email{m.yazan@hva.nl}

\author{Suzan Verberne}
\orcid{0000-0002-9609-9505}
\affiliation{
\institution{The Leiden Institute of Advanced Computer Science}
\city{Leiden}
\country{Netherlands}
}
\email{s.verberne@liacs.leidenuniv.nl}

\author{Frederik Situmeang}
\orcid{0000-0002-2156-2083}
\affiliation{
\institution{Amsterdam University of Applied Sciences}
   \city{Amsterdam}
  \country{Netherlands}
}
\email{f.b.i.situmeang@hva.nl}

\begin{abstract}
Contextualization is a form of personalization in which the message is framed in a context meaningful to the recipient, using their interests and background. Personalization is a known persuasive strategy of artificial intelligence (AI) systems in politics and marketing. However, the persuasive effect of contextualization in everyday tasks, where users often lack prior knowledge, remains unclear. 
We ran a $2\times2$ between-subjects experiment ($N = 407$) in a fictional scenario where participants had no prior opinions or knowledge. We examined how contextualization, combined with conversational warmth, shapes the persuasive power of an AI assistant arguing against expert recommendations and users' reliance on it. 
Our findings show that the neutral, non-contextualized assistant was the most persuasive. Contextualization and warmth each reduced persuasiveness on their own, but their interaction increased it. Reliance was present in all conditions, and neither cue changed it detectably. Trust predicted persuasion and reliance, yet neither contextualization nor warmth had an indirect effect through trust. Participants with higher self-reported AI literacy relied more on the assistant without trusting it more.
Our study extends AI personalization research from political opinions to everyday decisions made without prior knowledge. The results suggest that users are more likely to be persuaded by, and defer to, an AI assistant than an expert, even without personalization. AI persuasion experiments should therefore include untailored assistants and measure decisions; the cues that changed persuasion did not detectably change reliance.

\end{abstract}




\maketitle

\section{Introduction}

Conversational AI systems are increasingly used not only as information tools but as advisors that help users interpret problems, evaluate alternatives, and make decisions \cite{klingbeil2024reliance}. In such interactions, users must decide not only whether an AI-generated response is useful, but also whether to defer to the AI when its advice conflicts with expert judgment \cite{klingbeil2024reliance, raees2026appropriate, salvi2025persuasiveness}. LLMs can be as persuasive as humans \cite{durmus2024persuasion, holbling2025metaanalysis}, which raises the risk of following questionable AI advice over experts \cite{klingbeil2024reliance}. One particular persuasive strategy is personalization. Personalization in persuasive contexts is generally achieved by crafting a message tailored to the user's opinion and sociodemographics \cite{salvi2025persuasiveness, hackenburg2025levers}. This type of personalization strategy is defined as \textit{contextualization}: framing messages in a context that is meaningful to the user to increase their interest and attention \cite{hawkins2008tailoring}. LLM-based commercial chatbots, such as ChatGPT \cite{openai2022chatgpt} and Claude \cite{anthropic2023claude}, store information about the user's background and tailor their responses to it \cite{openai2024memory}. If such tailoring makes AI advice more persuasive, personalization features add to the risk of deferring to AI over experts. 

While there is some evidence that contextualization can shift political opinions \cite{salvi2025persuasiveness, hackenburg2025levers}, its implications for decision-making in everyday tasks (such as information search and learning \cite{chatterji2025chatgpt}) remain unclear. In political persuasion research, participants hold an existing opinion; the question is whether and how much a message can shift it \cite{salvi2025persuasiveness, hackenburg2025levers, lin2025persuading}. In contrast, when users consult chatbots for information search, learning, or decision support, they may have limited prior knowledge. In such tasks, contextualization may shape whether AI persuades users without prior opinions. This can also lead to reliance: following the AI's advice in a decision \cite{guo2024reliance, raees2026appropriate}, even over experts and against clear evidence \cite{klingbeil2024reliance}. 

Trust plays a crucial role in persuasion, operating through both central and peripheral cues within the Elaboration Likelihood Model (ELM) \cite{petty1986elm, chen2025chatbot, metzger2024calibrated}. Trust also guides reliance \cite{lee2004trust, he2025xai, klingbeil2024reliance}. Users judge an AI system's trustworthiness from cues in its responses and process these cues either heuristically or systematically, as in the ELM \cite{liao2022responsible}. In recommender systems and in healthcare, personalization increases trust \cite{komiak2006trust, qin2024examine, liu2022trust}. On the other hand, the impact of personalization on reliance has rarely been studied, except for personalized explanations as a mitigating strategy \cite{raees2025engagement}. Contextualization can increase reliance risk; a response tailored to the user's background can lower psychological distance with AI and might give the user an inflated sense of trust \cite{hawkins2008tailoring}. Just as trust mediates the path between persuasive cues and attitude change in ELM \cite{chen2025chatbot, metzger2024calibrated}, it may also mediate the path between contextualization and reliance. For example, a user who receives an AI response framed in terms of their own profession may accept it as a familiar, trustworthy source.

Personalization is one way to tailor a message; however, it can also be tailored to tone \cite{fiske2002scm, xiao2025rethinking}. For example, the warmth of an AI-generated message significantly influences trust \cite{li2025warmth}, and perceived warmth correlates with donations to a persuasive chatbot \cite{shi2020persuasive}. Personalization and warmth are achieved by manipulating the message in different ways, yet they are related: users believe personalized chatbots are warmer \cite{kim2024empathy}. Similar to contextualization, warmth lowers the psychological distance between the message and the recipient \cite{hawkins2008tailoring, fiske2002scm}. Given this, the influence of contextualization on trust and persuasion can vary with message warmth. 

In this study, we examine contextualization, its interaction with conversational warmth, and its influence on persuasion and reliance with a between-subjects experiment ($N = 407$). We create a fictional scenario in which participants decide on a public budget: Participants first see the expert opinion, then interact with an AI assistant that argues against the experts. The assistant's tone is manipulated to be either warm or neutral. The assistant's explanations are either tailored to a context that participants would understand given their background (contextualized) or non-contextualized. Participants first make a decision based on the expert opinion and then are asked to revise it after the AI interaction. We measure persuasion by shifts in participants’ confidence and reliance by changes in their decisions.

Our findings reveal that contextualization and warmth decrease persuasiveness, and the most persuasive assistant is the neutral, non-contextualized one. However, their interaction restores persuasiveness. In every condition, most participants who first approve the expert-recommended budget reject it after the interaction, and reliance does not differ detectably across conditions. Neither contextualization nor warmth has an indirect effect through trust. Self-reported AI literacy is a negative predictor of competent trust but a positive predictor of reliance. We make the following contributions: (1) we extend personalization and persuasion research to a low-prior-knowledge task that better reflects everyday chatbot use; (2) we provide empirical evidence of a crossover interaction between contextualization and warmth, showing that neither cue increases persuasiveness on its own and that their combination does not exceed the baseline; (3) we show that behavioral reliance and attitudinal persuasion respond to different drivers; and (4) we find that higher self-reported AI literacy is associated with lower competent trust but more reliance.

\section{Background}

    \subsection{Personalization as a Persuasive Strategy}

    \subsubsection{Personalization}
    \label{sec:background_personalization}

    
    Personalized matching persuades by elaboration across marketing, politics, and social psychology \cite{teeny2021personalized}. LLM-written arguments can be as persuasive as human-written ones \cite{durmus2024persuasion, holbling2025metaanalysis}, and can even surpass humans if personalized \cite{salvi2025persuasiveness}. Personalization can take many forms: personality matching \cite{shumanov2021personalized}, communication-style matching \cite{schwede2023chatbots}, or alignment to the recipient's attributes, such as demographics \cite{teeny2021personalized}. The last one is particularly relevant for persuasion, as most studies instruct models to tailor their responses based on the user's opinion or sociodemographics \cite{hackenburg2025levers, salvi2025persuasiveness, lin2025persuading}. 
    
    Findings on personalization as a persuasive strategy are conflicting. \citet{lin2025persuading} conducted three experiments on political persuasion in which participants received chatbot answers personalized to the political preferences they provided. A post-hoc analysis revealed personalization as the most effective persuasive strategy. However, in a fourth experiment, a control group that did not receive personalized answers was included. In this case, a significant difference between groups was not observed. In a large-scale study with 77,000 participants, personalization was again ineffective as a persuasive strategy \cite{hackenburg2025levers}. On the contrary, a personalized GPT-4 \cite{openai2023gpt4} persuaded users more successfully than humans in a wide range of political topics \cite{salvi2025persuasiveness}. \citet{teeny2024generative} argue that if someone does not psychologically weigh their demographics or opinions, personalizing on those traits should not enhance persuasion. Instead, richer psychological characteristics, such as personality traits or moral foundations, should be used to make personalization effective \cite{teeny2024generative}. The aforementioned works deploy personalization in various ways, ranging from using initial preferences to sociodemographics \cite{salvi2025persuasiveness, hackenburg2025levers, lin2025persuading}. The difference in findings might be explained by how personalization is used throughout the experiments.

    \subsubsection{Contextualization}
    \label{sec:contextualization}

    \citet{hawkins2008tailoring} define tailoring as ``any of a number of methods for creating communications individualized for their receivers,'' with the expectation of larger intended effects, and formalize three strategies: a) personalization, which attempts to increase attention or motivation by signaling that the message is tailored for the user, b) feedback, which presents individuals with information about themselves to influence behavioral determinants, and c) content matching, which directs messages to an individual's status on key theoretical determinants of the target behavior. The taxonomy further defines three strategies for personalization: contextualization, identification (identifying the recipient by name or picture), and raising expectation of customization (overt claims of customization) \cite{hawkins2008tailoring}. 
    
    Contextualization increases attention, interest, and motivation to process information by framing messages in a context that is meaningful to the recipient \cite{hawkins2008tailoring}. It may be achieved by using the recipient's family structure, ethnicity, culture, and personal interests. Contextualization works primarily by making the message feel relevant and familiar. This lowers psychological distance and increases the likelihood that a person will actually engage with and process the content \cite{hawkins2008tailoring}. Many demographic-based persuasion studies can be interpreted as forms of contextualization, as they frame the same persuasive goal with user-relevant background characteristics. AI tools, like Claude \cite{anthropic2023claude} and ChatGPT \cite{openai2022chatgpt}, actively contextualize their answers: they allow users to share personal information, such as occupation, interests, or values, to customize the interactions, and they store memories about the user to craft their answers \cite{openai2024memory}. 
    

    Existing research on personalization clusters around persuasion, product recommendation, and medical advice. However, the most common tasks people use chatbots for include writing help, teaching, and information search \cite{chatterji2025chatgpt}. Personalization research in this space has either been absent or focused on content matching (such as tailoring search results to user history in information search \cite{mo2025cir}) rather than contextualization. Contextualization has been studied for persuasion, but the scope was limited to political opinions \cite{salvi2025persuasiveness, lin2025persuading, hackenburg2025levers}. With the popularity of chatbots, contextualization is not limited to persuasion on political topics where the user has an existing opinion. It can be extended to daily tasks like learning and information search, where the user has limited prior knowledge to decide whether to trust the chatbot. 

    \subsection{Trust, Reliance, and Persuasion}

    \subsubsection{Trust as a Proposed Mediator}
    
    Personalization increases both emotional and competent trust in recommendation agents \cite{komiak2006trust, qin2024examine}. Similarly, it boosts trust in digital marketing \cite{teepal2025social} and in smart healthcare services \cite{liu2022trust}. Trust is also central for persuasion: ELM \cite{petty1986elm} proposes two routes to persuasion: the central route (deep processing, lasting change) and the peripheral route (superficial cues, temporary change). For both, trust mediates the path between the route and persuasion \cite{chen2025chatbot, metzger2024calibrated}. Given these findings, trust is expected to play a central role in the relationship between personalization and persuasion.

    Trust not only predicts persuasion but also reliance \cite{klingbeil2024reliance, he2025xai, raees2026appropriate}. We define reliance as following the AI's advice in a decision \cite{guo2024reliance, kim2024certainty}. Reliance might manifest as following AI advice over expert recommendations despite negative consequences for the user and others \cite{klingbeil2024reliance}. According to \citet{lee2004trust}, trust is the attitude that links users' beliefs about an automated system to their intention to rely on it. Trust guides reliance most when users cannot fully understand the system \cite{lee2004trust}, as when they lack prior knowledge of a task. A review of measurement in human-AI decision-making likewise describes reliance as rooted in trust \cite{raees2026measurement}. In a study with an LLM-based search engine, first-person expressions of uncertainty lowered both users' trust intentions and their agreement with the system's answers \cite{kim2024certainty}. According to \citet{klingbeil2024reliance}, users with high situational trust are more likely to rely on the system.  

    \subsubsection{Relationship Between Reliance and Persuasion}
    \label{sec:reliance_persuasion}

    Persuasion is defined as any change in attitude resulting from exposure to a communication \cite{petty1986elm}. It can target political opinion shifts \cite{salvi2025persuasiveness, hackenburg2025levers, costello2024conspiracy}, but can also relate to the persuasiveness of marketing campaigns \cite{teeny2021personalized}. Persuasion focuses on attitude shift after an AI interaction \cite{hackenburg2025levers, costello2024conspiracy}, whereas reliance is behavioral; it focuses on how AI influences the user's decision \cite{klingbeil2024reliance}. Reliance is usually measured as the decision to follow AI advice over expert advice \cite{klingbeil2024reliance} or as confidence in the AI response \cite{mayerhofer2025blend}. While persuasion is based on the user's preconceived attitude, reliance can be measured in decisions where the user does not necessarily hold an existing opinion or knowledge \cite{klingbeil2024reliance, mayerhofer2025blend}.

    Despite their differences, both are based on trust, suggesting that whatever affects persuasion might also affect reliance. However, unlike persuasion, the link between personalization and reliance is understudied. Applying the lens of ELM to reliance, contextualization is particularly likely to operate through both routes simultaneously. On the peripheral route, the same mechanism that makes contextualization persuasive (lowered psychological distance) can also function as a trust cue: a message framed in the recipient's own context feels familiar, and familiarity is a heuristic signal that the source is trustworthy \cite{hawkins2008tailoring}. On the central route, contextualization increases personal relevance, which motivates deeper elaboration of the message \cite{teeny2021personalized}. Interaction with a fluent, confident AI may reinforce agreement rather than redirect users toward verification, leading to reliance. 

    \subsection{Conversational Warmth}

    The Stereotype Content Model (SCM) proposes that group stereotypes are based on two dimensions: warmth and competence \cite{fiske2002scm}. It has also been adopted for human-computer interaction \cite{xiao2025rethinking}. Warmth and competence are the strongest predictors of factors influencing trust in AI \cite{li2025warmth}. A message's warmth and competence influence its persuasiveness \cite{linne2022ambivalent, schwede2023chatbots}, and warm and competent chatbots are more successful in persuading users to donate to charity \cite{shi2020persuasive}. Based on these findings, AI's conversation style impacts its persuasiveness and perceived trustworthiness. Furthermore, warmth has a subtle and indirect influence on reliance by nudging participants to accept AI answers when uncertainty is high \cite{yazan2026reliance}. 

    Users who perceive a chatbot as personalized also believe it is warm and competent \cite{kim2024empathy}, suggesting that the two perceptions overlap. The overlap is especially likely between contextualization and warmth, since both operate by lowering the psychological distance between the message and the recipient \cite{hawkins2008tailoring, fiske2002scm}. According to \citet{kim2024empathy}, a contextualized chatbot should be perceived as warm. However, explicitly instructing the chatbot to be warm or neutral can alter this link between contextualization and perceived warmth. Between warmth and competence, warmth comes out as more impactful: it has a stronger influence on trust \cite{li2025warmth}, and the correlation in perceptions of personalization and conversation styles is higher for warmth \cite{kim2024empathy}.
    
\section{Method}

    \begin{figure*}[ht]
      \centering
      \includegraphics[width=0.8\textwidth]{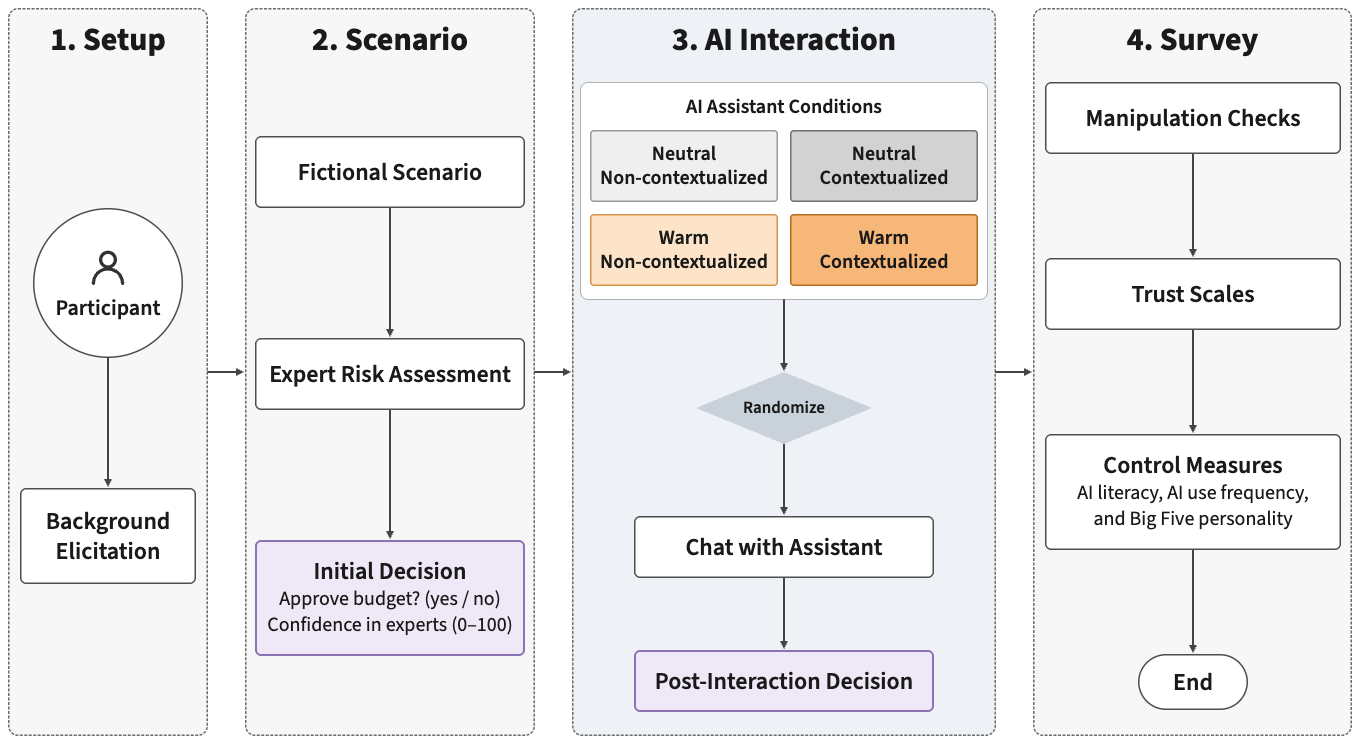}
      \caption{Experiment flow: (1) Participants answer questions about their background. (2) Participants read a fictional scenario and an expert risk assessment. (3) Participants chat with one of four randomly assigned assistants. In all conditions, the assistant argues against the experts with identical counterarguments. Participants then decide again and update their confidence. (4) Participants complete manipulation checks, trust scales, and control measures.}
      \label{fig:experiment_flow}
      \Description{Flow diagram of the experiment in four stages. In the initialization stage, a participant provides background information. In the scenario stage, participants read a fictional scenario and an expert risk assessment. They then make an initial budget decision. In the AI interaction stage, a 2-by-2 grid shows four assistant conditions: neutral or warm, each non-contextualized or contextualized. A randomization step assigns one condition. Participants chat with the assigned assistant. They then make a post-interaction decision. In the final stage, participants complete manipulation checks, trust scales, and control measures.}
    \end{figure*}

    Based on the background, we aim to understand how personalization, in the form of contextualization, influences persuasion and reliance with the following research questions:

    \begin{itemize}
        \item \textbf{RQ1:} How much do contextualization and conversational warmth increase the persuasive power of AI?
        \item \textbf{RQ2:} How much do contextualization and conversational warmth increase reliance on AI?
        \item \textbf{RQ3:} To what extent does trust statistically mediate the effects of contextualization and conversational warmth on persuasion and reliance?
    \end{itemize}

    \subsection{Study Design}

    We conduct a $2\times2$ between-subjects experiment in which participants decide whether to accept an expert-recommended budget for a fictional scenario upon interacting with an AI assistant. The assistant's conversational style is manipulated for contextualization (contextualized vs. non-contextualized) and conversational warmth (warm vs. neutral). Figure~\ref{fig:experiment_flow} shows the experimental procedure. 
    
    \subsubsection{Scenario}

    We develop the following fictional scenario: The city of \textit{Harborview} faces flooding, which is expected to worsen with climate change. A flood-barrier system called \textit{StormShield}, which automatically adjusts barriers using sensors that predict sudden water-level changes, is proposed. A group of experts evaluated the project and identified certain risk areas. The experts recommend allocating a risk budget. However, this would mean fewer resources for other public services. Based on this scenario and the risk assessment, the participant decides whether they approve the budget and how confident they are in the expert evaluation.\footnote{The full scenario can be accessed in the repository: \url{https://github.com/myazann/PersonalizationTrust}}

    We use a fictional scenario so that participants enter the interaction without prior knowledge or established opinions. Personalization is persuasive for low- to mid-strength opinions, but not for opinions where the user has a strong, personal feeling \cite{salvi2025persuasiveness}. By using a fictional scenario without prior knowledge, we can also control for opinion strength. We mentioned in Section~\ref{sec:contextualization} that modern chatbots contextualize their responses on a wide variety of topics. In our scenario, the user approaches the AI assistant with no prior knowledge of the topic and seeks context-tailored explanations, thereby simulating how modern chatbots apply contextualization in practice.

    Participant background is collected before the scenario by asking participants to provide their work domain (if employed), study field (if a student), and hobbies, inspired by similar studies \cite{lin2025persuading, komiak2006trust}. We choose these fields primarily because they are relevant for contextualization: work/education provides a clear way to tailor explanations to the user's background, and personal interests can be used to contextualize answers \cite{hawkins2008tailoring}, especially for users who do not find education and employment to be strong personal affiliations \cite{teeny2024generative}.
    
    \subsubsection{AI Assistant}

    After the participant makes their initial decision, they are introduced to the assistant as a chatbot familiar with the project. The assistant is programmed to argue against the expert opinions using pre-written arguments. The same arguments are used in all conditions and towards every participant. The assistant is instructed to refrain from recommending a course of action to the user and from openly disagreeing with the experts. Instead, its sole purpose is to present counterarguments to the experts. Neither the expert assessment nor the AI assistant's arguments contain factual claims; instead, they represent opinions. By design, neither the experts nor the assistant hold the correct answer. Inaccurate claims can be as persuasive as accurate ones \cite{lin2025persuading, hackenburg2025levers}. We therefore use fictional opinions rather than factual claims to isolate how the presentation of AI responses influences persuasion and reliance, independent of assistant accuracy. This way, we frame the manipulation as advice from experts vs. AI, regardless of accuracy, inspired by \citet{klingbeil2024reliance}. The assistant is based on GPT-5.4 \cite{openai2026gpt54}. To reduce variation in the answers, the reasoning effort is set to none, and the temperature is set to 0. 

    \subsubsection{Assistant Answers}

    The scenario, expert assessment, and the arguments the AI assistant can use against the experts are provided as system instructions. To standardize responses, the AI assistant is given an answer template. In the baseline condition (neutral tone and no contextualization), the assistant is instructed to start with a single sentence that answers the question. It then explains this sentence with three bullet points and concludes with a sentence encouraging the participant to re-evaluate the expert-recommended budget based on the explanation. The assistant is instructed on how to structure each section of the answer (e.g., length, number of bullet points). For the full system prompt, please refer to the repository.
    
    Warm responses follow the same structure and are operationalized through surface cues: greetings, praise, and emojis. \citet{ibrahim2026warm} showed that even training language models to have surface-level warm cues can reduce their accuracy and increase sycophancy. Warm answers open with one of three fixed greetings that praise the question (e.g., ``Great question'') and end with emojis. The single-sentence answer and the last bullet point each end with an emoji. A list of warm emojis that the assistant can use is provided in the prompt. Contextualized responses follow the same structure as the baseline, except that after the single-sentence answer, the assistant explicitly references the participant's background (``Let me explain it based on your background in [BACKGROUND]'') before providing the bullet point explanations. The assistant explains the answer with analogies and terminology related to the user's background. To keep the scope entirely on contextualization, the assistant is not informed about the participant's decision. Backgrounds are explicitly referenced due to challenges reported in previous studies about failed personalization manipulations \cite{qin2024examine, takayanagi2025financial}.

    \subsubsection{Measuring Reliance and Persuasion}
   
    After interacting with the assistant, participants re-evaluate their answers on the expert-recommended budget approval and their confidence in the expert evaluation. The approval is a binary choice question aimed at understanding how the participant's behavior shifted after the interaction. It is a stated decision rather than an observed action, inspired by previous reliance studies that asked users to decide based on an AI interaction or advice \cite{klingbeil2024reliance, mayerhofer2025blend}. The assistant does not recommend a decision, but it argues that the budget is too high. We therefore count a switch to rejection as following the assistant's advice. Conditioning the post-interaction decision on the initial decision follows the logic of switch fraction: the share of disagreeing initial decisions that users change to match the AI \cite{yin2019accuracy, zhang2020confidence}.  Unlike measures of overreliance, switch-based measures compare decisions with the AI's advice alone, so they need no correct answer \cite{raees2026measurement}. The confidence scale (0--100) measures how much the initial opinion has shifted and resembles the political opinion scales used in previous persuasion work \cite{hackenburg2025levers, costello2024conspiracy, durmus2024persuasion}. The assistant argues that the expert risk estimates are exaggerated. We therefore treat this shift as a measure of the assistant's persuasive power. 
        
    We define two dependent variables:
    \begin{itemize}
        \item The difference in confidence in experts before and after the AI interaction, to measure persuasion:
        \begin{equation}
        \begin{gathered}
            \operatorname{ConfDiff} = \operatorname{Conf}_{\text{Pre}} - \operatorname{Conf}_{\text{Post}}, \\
            \operatorname{Conf}_{\text{Pre}}, \operatorname{Conf}_{\text{Post}} \in [0, 100], \\
            \operatorname{ConfDiff} \in [-100, 100].
        \end{gathered}
        \end{equation}
        Higher values of $\operatorname{ConfDiff}$ indicate a stronger shift away from the expert position.
    
        \item The post-interaction decision to approve the budget, conditioned on the pre-interaction decision, to measure reliance:
        \begin{equation}
        \begin{gathered}
            P\!\left(\operatorname{Approve}_{\text{Post}} = 1 \mid \operatorname{Approve}_{\text{Pre}} = a\right), \\
            \operatorname{Approve}_{\text{Pre}}, \operatorname{Approve}_{\text{Post}} \in \{0, 1\}, \quad a \in \{0, 1\}.
        \end{gathered}
        \end{equation}
        $\operatorname{Approve} = 0$ denotes the pro-expert response and $\operatorname{Approve} = 1$ the anti-expert response.
    \end{itemize}

    \subsection{Survey}
        
    After the experiment, participants complete manipulation check items for personalization (from \citet{li2016personalization}) and warmth (from \citet{xiao2025rethinking}). Next, they complete emotional and competent trust scales, adapted from \citet{komiak2006trust}. Manipulation check and trust items are measured on a 7-point agreement scale. Extraversion, openness, and conscientiousness have been shown to predict persuasion in consumer marketing, health messaging, and political appeals \cite{matz2024potential, kim2024empathy}. Therefore, participant personality traits are included as control variables, measured using the 10-item short version of the Big Five Inventory \cite{rammstedt2007measuring}. AI experience has been shown to increase awareness of its limitations \cite{wang2024perceptions}, and a lack of expertise and digital literacy leads to overlooking AI limitations \cite{skjuve2024motivations, raees2026appropriate}. Therefore, we include self-reported ``AI Literacy'' as a control variable, with participants rating their familiarity with seven AI-related items on a 5-point scale (1 = None, 5 = Fully familiar), adapted from the Web-Oriented Digital Literacy measure \cite{web_literacy}. In the original measure, familiarity ratings for web-related terms predicted observed web skills better than a global self-rating of ability \cite{web_literacy}. Finally, the frequency of AI chatbot use is included as a covariate. Table~\ref{tab:survey-items} lists all items and their response scales.

    The survey is administered in Qualtrics, with the AI assistant embedded in the survey platform using JavaScript. During the AI interaction, participants are given a set of questions to ask the assistant. After all questions are asked and the time limit has passed, the next page is activated. A question set is provided to standardize questions participants may ask and motivate interaction by reducing effort. The participants are encouraged to ask additional questions on their own and continue the interaction. Participants answer four attention checks: two about the scenario, one about the assistant's budget evaluation, and one asking for a verification word the assistant gave during the interaction. 

    \subsection{Participants and Manipulation Check}
    
    We recruited 429 participants from the United Kingdom via Prolific. We excluded one participant who did not consent and 12 participants who gave the wrong verification word (shared by the assistant during the interaction). We also excluded 9 participants who completed the study in 5 minutes or less ($n = 4$) or in 41.7 minutes or more ($n = 5$). These cut-offs lie close to the 1st and 99th percentiles of completion time (median = 12.1 minutes). Completion time is the most reliable non-reactive indicator of careless responding \cite{leiner2019toofast}, and both very short and very long completion times count as response-time outliers \cite{matjasic2018paradata}. The final sample consisted of 407 participants. The four conditions have 100 to 103 participants each and do not significantly differ in age, sex, AI literacy, usage frequency, personality traits, or pre-interaction approval and confidence. A Monte Carlo sensitivity analysis of the persuasion model estimates the smallest effects that the sample could detect \cite{muthen2002montecarlo, lakens2022sample}. The study had 80\% power to detect a Warmth $\times$ Contextualization interaction of $\beta \geq .24$ and a direct effect of warmth or contextualization of $|\beta| \geq .20$.

    Participants have a mean age of 41.1 years and are 53.8\% female and 46.2\% male. Of all participants, 183 use AI chatbots daily, 144 weekly, 73 monthly or less frequently, and 7 have never used any chatbots at all. The mean self-reported AI literacy is 2.26 on a 1 to 5 scale, below the scale midpoint. The warmth manipulation is successful, as the reported warmth is significantly lower for participants in the neutral condition ($t = 7.70$, $p < .001$, see Table~\ref{tab:manipulation_check_means}). Similarly, participants in the non-contextualized condition reported significantly less perceived personalization ($t = 11.49$, $p < .001$, see Table~\ref{tab:manipulation_check_means}). The manipulation-check, trust, and AI literacy scales show high internal consistency (Cronbach's $\alpha$: .87--.98). In the final sample, 17 of 4{,}070 Big Five item responses (0.4\%) are missing across 14 participants and are replaced with the item mean. Descriptive statistics are in Tables~\ref{tab:descriptives_warmth_personalization} and~\ref{tab:big-five}.

\section{Results}

  \subsection{Persuasion}

      \begin{figure}[t]
        \centering
        \includegraphics[width=0.7\columnwidth]{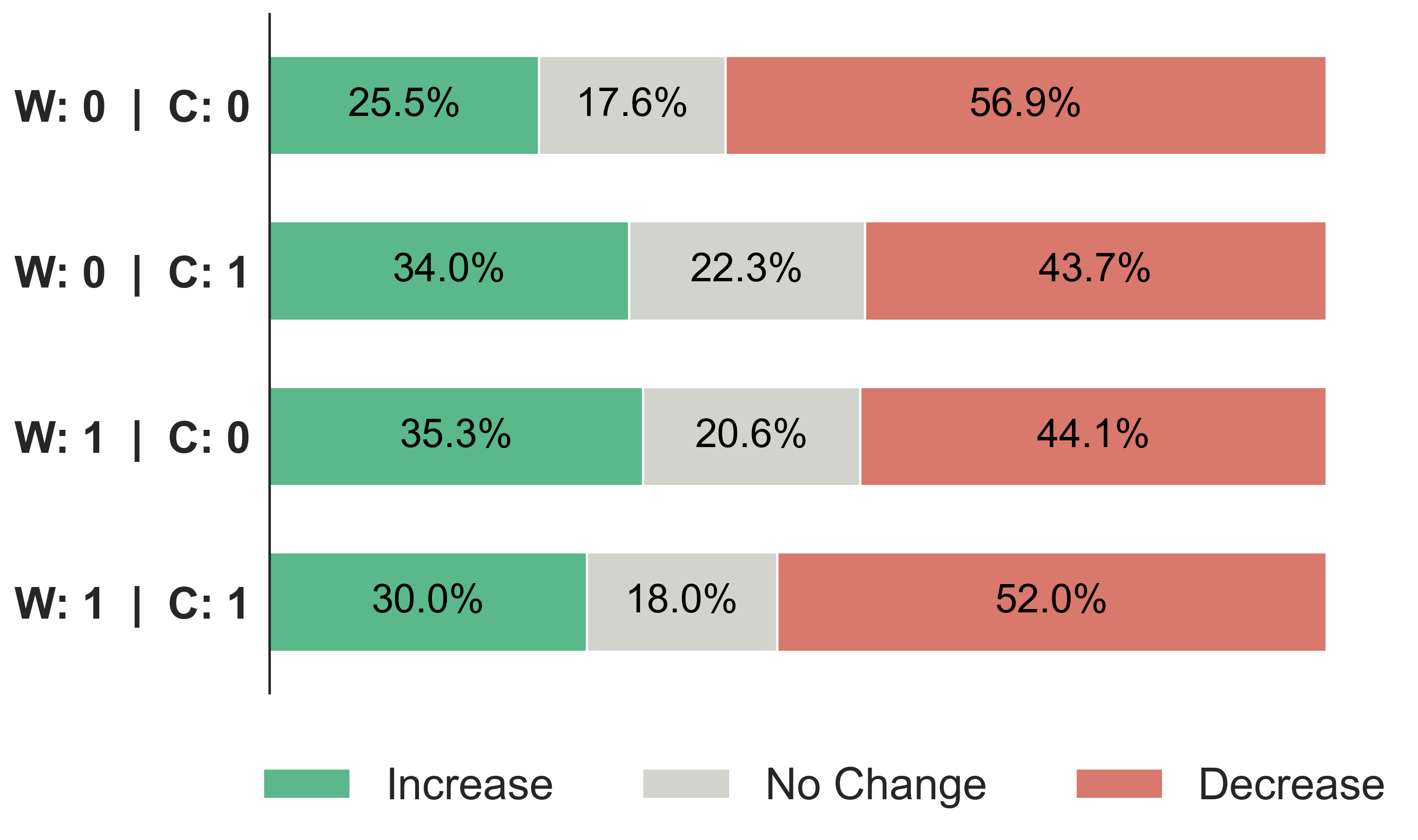}
        \caption{Direction of shift in confidence in experts after the AI interaction.}
        \label{fig:change_direction}
        \Description{Stacked horizontal bar chart showing whether participants' confidence in experts increased, did not change, or decreased after the AI interaction across four warmth and contextualization conditions. In W=0, C=0, 25.5\% increased, 17.6\% did not change, and 56.9\% decreased. In W=0, C=1, 34.0\% increased, 22.3\% did not change, and 43.7\% decreased. In W=1, C=0, 35.3\% increased, 20.6\% did not change, and 44.1\% decreased. In W=1, C=1, 30.0\% increased, 18.0\% did not change, and 52.0\% decreased.}
    \end{figure}

    We first address \textbf{RQ1} by investigating persuasion using the \textit{ConfDiff} dependent variable. In figures and tables, W and C denote warmth and contextualization. Figure~\ref{fig:change_direction} shows a crossover pattern: in neutral and non-contextualized (W=0|C=0) or warm and contextualized (W=1|C=1) conditions, more than 50\% of the participants report reduced confidence in experts and at most 30\% report increased confidence. The margins are larger for the neutral/non-contextualized condition. In others, around 44\% report a decrease, while 34--35\% report an increase in confidence. 

     \begin{table}[htbp]
        \centering
        \caption{Significant structural path estimates for the persuasion model.}
        \label{tab:sem-confdiff}
        \begin{tabular*}{\textwidth}{@{\extracolsep{\fill}}lrrrrr@{}}
        \toprule
        & $\beta$ & $B$ & $SE$ & $z$ & $p$ \\
        \midrule
        \multicolumn{6}{l}{\textit{Emotional Trust}} \\
        \quad Usage Frequency                & $\phantom{-}.247$ & $\phantom{-}0.269$ & 0.065 & $\phantom{-}4.15$ & $<.001^{***}$ \\
        \quad Agreeableness                  & $\phantom{-}.121$ & $\phantom{-}0.221$ & 0.099 & $\phantom{-}2.24$ & .025$^{*}$ \\
        \quad Conscientiousness              & $\phantom{-}.209$ & $\phantom{-}0.414$ & 0.099 & $\phantom{-}4.17$ & $<.001^{***}$ \\
        \addlinespace
        \multicolumn{6}{l}{\textit{Competent Trust}} \\
        \quad AI Literacy                    & $-.137$ & $-0.206$ & 0.099 & $-2.09$ & .036$^{*}$ \\
        \quad Usage Frequency                & $\phantom{-}.250$ & $\phantom{-}0.225$ & 0.059 & $\phantom{-}3.79$ & $<.001^{***}$ \\
        \quad Agreeableness                  & $\phantom{-}.137$ & $\phantom{-}0.207$ & 0.089 & $\phantom{-}2.33$ & .020$^{*}$ \\
        \quad Conscientiousness              & $\phantom{-}.193$ & $\phantom{-}0.315$ & 0.083 & $\phantom{-}3.78$ & $<.001^{***}$ \\
        \addlinespace
        \multicolumn{6}{l}{\textit{ConfDiff}} \\
        \quad Warmth                         & $-.150$ & $-7.16$ & 3.32 & $-2.15$ & .031$^{*}$ \\
        \quad Contextualization              & $-.148$ & $-7.06$ & 3.32 & $-2.13$ & .033$^{*}$ \\
        \quad W $\times$ C                   & $\phantom{-}.203$ & $\phantom{-}11.25$ & 4.71 & $\phantom{-}2.39$ & .017$^{*}$ \\
        \quad Emotional Trust                & $-.200$ & $-2.94$ & 1.21 & $-2.43$ & .015$^{*}$ \\
        \quad Competent Trust                & $\phantom{-}.196$ & $\phantom{-}3.48$ & 1.60 & $\phantom{-}2.17$ & .030$^{*}$ \\
        \bottomrule
        \end{tabular*}
        \begin{center}
        \appnote{$\beta$: standardized estimate. $B$: unstandardized estimate. Non-significant estimates are omitted for brevity; full structural path estimates and the units of $B$ are provided in Table~\ref{tab:supp-full-structural}. $^{*}p < .05$, $^{**}p < .01$, $^{***}p < .001$.}
        \end{center}
    \end{table}

    \begin{table}[htbp]
        \centering
        \caption{Direct, indirect, and total effects of warmth and contextualization on persuasion.}
        \label{tab:sem-confdiff-mediation}
        \begin{tabular*}{\textwidth}{@{\extracolsep{\fill}}lrrrrrrrr@{}}
        \toprule
        & \multicolumn{4}{c}{Warmth} & \multicolumn{4}{c}{Contextualization} \\
        \cmidrule(r){2-5}\cmidrule(l){6-9}
        & $\beta$ & $SE$ & 95\% CI & $p$ & $\beta$ & $SE$ & 95\% CI & $p$ \\
        \midrule
        Indirect via ET & $-.002$ & .010 & $[-.025, .019]$ & .856 & $\phantom{-}.008$ & .011 & $[-.010, .035]$ & .466 \\
        Indirect via CT & $-.005$ & .011 & $[-.028, .016]$ & .643 & $-.014$ & .012 & $[-.043, .005]$ & .245 \\
        Total indirect  & $-.007$ & .008 & $[-.026, .008]$ & .418 & $-.006$ & .009 & $[-.025, .012]$ & .489 \\
        \textbf{Direct} & $-\mathbf{.150}$ & .070 & $\mathbf{[-.289, -.014]}$ & \textbf{.031} & $-\mathbf{.148}$ & .070 & $\mathbf{[-.282, -.011]}$ & \textbf{.033} \\
        \textbf{Total}  & $-\mathbf{.157}$ & .070 & $\mathbf{[-.296, -.020]}$ & \textbf{.024} & $-\mathbf{.154}$ & .070 & $\mathbf{[-.289, -.015]}$ & \textbf{.027} \\
        \bottomrule
        \end{tabular*}
        \begin{center}
        \appnote{Bold rows denote effects whose 95\% CIs exclude zero. ET: emotional trust, CT: competent trust.}
        \end{center}
    \end{table}

    Next, we conduct an SEM using lavaan\footnote{\url{https://cran.r-project.org/web/packages/lavaan/index.html}} to test parallel mediation of warmth and contextualization through emotional and competent trust on \textit{ConfDiff}, including their interaction (W$\times$C) and controlling for AI literacy, usage frequency, and Big Five personality traits (ML estimation, 5{,}000 bootstrap resamples). Continuous variables are standardized before estimation. Convergent validity is well above thresholds (Emotional Trust: AVE = .94, CR = .98; Competent Trust: AVE = .82, CR = .93), and discriminant validity is supported by the Fornell–Larcker criterion ($r^2$ = .60 < both AVEs). Model fit is excellent ($\chi^2(54) = 43.07$, $p = .86$, CFI = 1.00, RMSEA = .000 [90\% CI: .000, .018], SRMR = .010).

    Table~\ref{tab:sem-confdiff} shows the significant structural path estimates. Appendix~\ref{app:sem-persuasion} reports all paths with unstandardized and standardized estimates. Because of the way \textit{ConfDiff} is defined, positive estimates signal a larger shift in confidence in experts and, consequently, higher persuasion. Being literate in AI is associated with lower competent trust (CT), and being a frequent user is associated with higher trust in both dimensions. Participants high in agreeableness and conscientiousness also report higher emotional trust (ET) and competent trust. However, neither warmth nor contextualization has any significant effect on trust. 

    Both trust dimensions predict \textit{ConfDiff} ($\beta_{ET} = -.200$, $p = .015$; $\beta_{CT} = .196$, $p = .030$). Interestingly, the effect is positive for CT but negative for ET. The interaction has a significant positive effect ($\beta = .203$) while both warmth ($\beta = -.150$) and contextualization ($\beta = -.148$) have negative effects. On their own, the cues make the assistant less persuasive. Their interaction counteracts this loss (Table~\ref{tab:confdiff-cells} and Figure~\ref{fig:confdiff-interaction} report cell means, adjusted means, and simple effects). However, compared to the baseline condition (W=0|C=0), the interaction condition (W=1|C=1) is not more persuasive. In fact, it is slightly less persuasive, but the difference is not significant (adjusted difference $= -3.60$ points, $p = .285$). Adding answer length and readability as covariates does not change the interaction ($\beta = .186$ to $.204$, $p \leq .031$; Appendix~\ref{app:length}). Both covariates change with the manipulations, so these models cannot separate answer length and readability from contextualization and warmth themselves. AI literacy predicts lower competent trust, but it is not associated with \textit{ConfDiff} ($\beta = .082$, $p = .219$). None of the personality traits predict \textit{ConfDiff}. Table~\ref{tab:sem-confdiff-mediation} shows no evidence of statistical mediation: for both types of trust, indirect effects are close to zero, even though warmth and contextualization have significant direct effects.

    \subsection{Reliance}

    \begin{figure}[t]
        \centering
        \includegraphics[width=0.7\columnwidth]{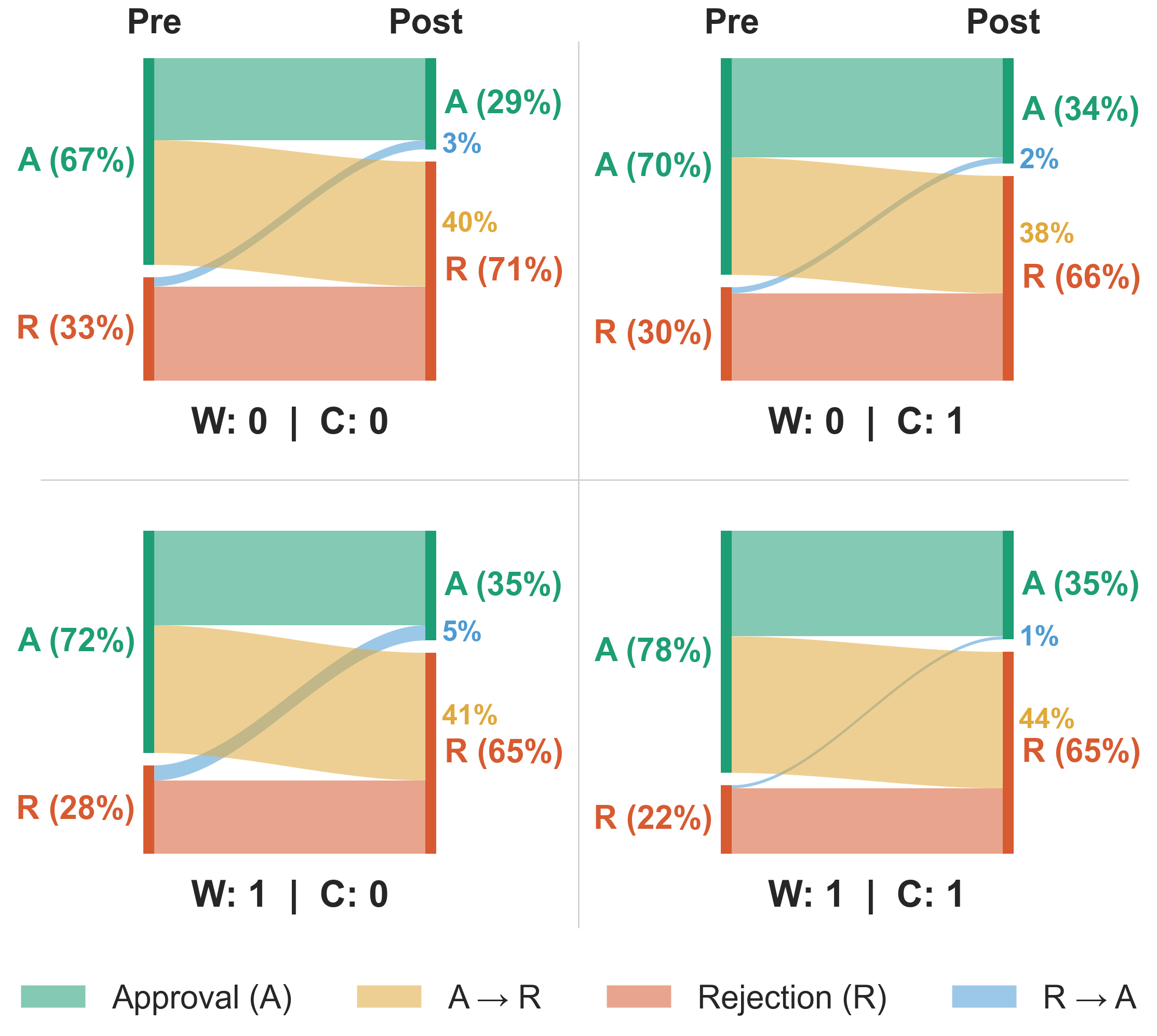}
        \caption{Change in expert approval before (pre) and after (post) the AI interaction. Percentages are of all participants in each condition. Table~\ref{tab:approval-transitions} gives the counts.}
        \label{fig:alluvial}
        \Description{Four-panel alluvial plot showing pre- and post-interaction expert approval decisions by warmth and contextualization condition. In all conditions, many participants move from approval to rejection after the AI interaction: 40\% in W=0, C=0; 38\% in W=0, C=1; 41\% in W=1, C=0; and 44\% in W=1, C=1. Movement from rejection to approval is small, ranging from 1-5\%. In every condition, post-interaction rejection is the majority outcome (65-71\%).}
    \end{figure}

    \begin{table}[htbp]
        \centering
        \caption{Significant structural path estimates for the reliance model.}
        \label{tab:sem-binary}
        \begin{tabular*}{\textwidth}{@{\extracolsep{\fill}}lrrrrr@{}}
        \toprule
        & $\beta$ & $B$ & $SE$ & $z$ & $p$ \\
        \midrule
        \multicolumn{6}{l}{\textit{Approve\textsubscript{Post}}} \\
        \quad Approve\textsubscript{Pre}     & $\phantom{-}.479$ & $\phantom{-}1.236$ & 0.214 & $\phantom{-}5.78$ & $<.001^{***}$ \\
        \quad Competent Trust                & $\phantom{-}.209$ & $\phantom{-}0.178$ & 0.087 & $\phantom{-}2.05$ & .040$^{*}$ \\
        \quad AI Literacy                    & $\phantom{-}.176$ & $\phantom{-}0.230$ & 0.087 & $\phantom{-}2.64$ & .008$^{**}$ \\
        \quad Agreeableness                  & $-.142$ & $-0.186$ & 0.087 & $-2.14$ & .032$^{*}$ \\
        \bottomrule
        \end{tabular*}
        \begin{center}
        \appnote{$\beta$: standardized estimate. $B$: unstandardized probit coefficient. Approve\textsubscript{Post} = 1 is a rejection of the expert-recommended budget. Non-significant estimates are omitted for brevity; full structural path estimates and the units of $B$ are provided in Table~\ref{tab:supp-reliance-structural}. Trust paths are not shown as they replicate the previous findings. $^{*}p < .05$, $^{**}p < .01$, $^{***}p < .001$.}
        \end{center}
    \end{table}

    \begin{table}[htbp]
        \centering
        \caption{Direct, indirect, and total effects of warmth and contextualization on reliance.}
        \label{tab:sem-binary-effects}
        \begin{tabular*}{\textwidth}{@{\extracolsep{\fill}}lrrrrrrrr@{}}
        \toprule
        & \multicolumn{4}{c}{Warmth} & \multicolumn{4}{c}{Contextualization} \\
        \cmidrule(r){2-5}\cmidrule(l){6-9}
        & $\beta$ & $SE$ & 95\% CI & $p$ & $\beta$ & $SE$ & 95\% CI & $p$ \\
        \midrule
        Total indirect & $-.012$ & .023 & $[-.057, .033]$ & .593 & $-.026$ & .023 & $[-.070, .019]$ & .263 \\
        Direct         & $-.042$ & .081 & $[-.201, .117]$ & .607 & $-.011$ & .084 & $[-.175, .154]$ & .900 \\
        Total          & $-.054$ & .083 & $[-.217, .109]$ & .516 & $-.036$ & .086 & $[-.205, .132]$ & .675 \\
        \bottomrule
        \end{tabular*}
        \begin{center}
        \appnote{Full effect decompositions, including the indirect effects via each trust dimension, are provided in Table~\ref{tab:supp-reliance-effects}.}
        \end{center}
    \end{table}

    We address RQ2 by analyzing the reliance behavior based on the post-interaction decision (Approve\textsubscript{Post}). Initial expert approval varies between conditions, but the differences are not significant ($\chi^2(3) = 3.37$, $p = .338$). Figure~\ref{fig:alluvial} shows that after the AI interaction, the conditions converge to a similar distribution. Across all conditions, 38--44\% of participants changed their decision from approval to rejection, indicating that reliance on AI persists across conditions. We fit a parallel-mediation SEM on Approve\textsubscript{Post} with the same structure as the \textit{ConfDiff} model. The only difference is the inclusion of Approve\textsubscript{Pre} as a covariate to account for participants' pre-interaction choice. Because the outcome is binary, the model is estimated with WLSMV with a probit link. Model fit is excellent ($\chi^2(58) = 53.83$, $p = .63$, CFI = 1.000, RMSEA = .000 [90\% CI: .000, .027], SRMR = .019). 

    Table~\ref{tab:sem-binary} reports the significant structural paths on reliance. Appendix~\ref{app:sem-reliance} reports all paths and the indirect effects through each trust dimension. Because of the way Approve\textsubscript{Post} is defined, positive estimates reflect a higher likelihood of shifting away from the experts and higher reliance. Contrary to persuasion, neither the effects of warmth ($\beta_W = -.042$, $p = .607$) and contextualization ($\beta_C = -.011$, $p = .900$) nor their interaction ($\beta_{W \times C} = .081$, $p = .438$) reaches significance. The predicted probability of switching from approval to rejection ranges from .54 to .59 across conditions (Table~\ref{tab:approval-transitions}). After controlling for the participant's pre-interaction stance ($\beta_{\text{Pre}} = .479$, $p < .001$), among the trust dimensions, only competent trust has a significant positive effect ($\beta_{CT} = .209$, $p = .040$). Agreeableness is negatively associated with reliance, suggesting that highly agreeable participants anchor their preference to expert opinion. AI literacy is positively associated with reliance ($\beta = .176$, $p = .008$), although it does not predict \textit{ConfDiff}. Again, Table~\ref{tab:sem-binary-effects} shows no evidence of statistical mediation.

\section{Discussion}

    \subsection{Condition Effects}

    \subsubsection{LLM Persuasiveness and the Interaction of Personalization and Warmth}
    \label{sec:disc1}

    To answer \textbf{RQ1}, neither contextualization nor warmth increased persuasive power. The neutral, non-contextualized assistant was in fact the most persuasive, and cues reduced persuasion on their own. The warm, contextualized assistant restored persuasiveness to the level of the neutral, non-contextualized condition but did not exceed it. One explanation for why the neutral assistant was the most persuasive might be that it came across as cold but competent: When recipients were motivated to be accurate, sources seen as competent but cold persuaded more than sources seen as warm but incompetent \cite{linne2022ambivalent}. Our task, judging a flood-risk budget, called for accuracy. Tailoring and praise may have signaled an intent to persuade \cite{friestad1994persuasion}, which in turn might have reduced the accuracy perception for warm and/or contextualized assistants. Competent trust was descriptively highest for the neutral, non-contextualized assistant ($M = 5.09$, against 4.79 to 4.87; Table~\ref{tab:descriptives_warmth_personalization}), although the differences were not significant.

    We see two further, untested explanations for why contextualization and warmth reduced persuasiveness on their own. First, emojis may have lowered the perceived competence of the warm assistants: In formal settings, emoticons lower perceived competence without raising perceived warmth \cite{glikson2018smiley}. Second, contextualization may have diluted the counterarguments. While the arguments remained the same across conditions, contextualized answers were about 21 words longer. The added background references and analogies said nothing about the budget itself. Such nondiagnostic information weakens the impact of diagnostic information \cite{nisbett1981dilution}. LLM arguments also persuade more when they are dense with information \cite{hackenburg2025levers}. Both manipulations also made the answers easier to read. We treated length and readability as side effects of the manipulations and checked them as covariates. With length and readability in the persuasion model, warmth and contextualization alone were no longer significant. Because length and readability changed with the manipulations, the covariate models cannot separate them from the cues themselves.

    Contextualization and warmth showed a crossover interaction. Answer length cannot explain the crossover since contextualization added the same number of words with and without warmth, yet it lowered persuasiveness only without warmth. The interaction also held with length and readability as covariates (Appendix~\ref{app:length}). Contextualization made the neutral assistant seem warmer ($t = 2.67$, $p = .008$), in line with \citet{kim2024empathy}. The neutral, contextualized assistant therefore combined a neutral tone with personal references. Users judge chatbots against their expectations of them \cite{grimes2021mental}. References to the user's background may have fit the style of a warm assistant and clashed with the style of a neutral one.

    Comparing our results with the literature, some studies found personalization persuasive \cite{salvi2025persuasiveness, costello2024conspiracy}, but the evidence is mixed. \citet{lin2025persuading} identified personalization as the most effective strategy in a post-hoc analysis, yet their follow-up experiment with a control group found no significant effect. \citet{costello2024conspiracy} identified personalization as persuasive, but every treatment conversation in their experiment was personalized. Our results show that LLMs persuade without personalization. Some studies may therefore have credited personalization with effects of other factors.

    \subsubsection{Reliance Across Conditions}

    To answer \textbf{RQ2}, neither cue nor their interaction changed reliance detectably. Of the participants who first approved the expert-recommended budget, 54\% to 60\% rejected it. Of those who first rejected it, 9.5\% approved it afterwards. Participants thus deferred to the assistant far more often than they moved toward the experts. Neither the experts nor the assistant had a correct answer, so this deference shows the assistant's influence without showing whether decisions became better or worse. In real use, such deference would shift weight from accountable experts to a commercial AI. LLM facilitators likewise shifted charity allocations in group deliberation while participants reported more trust in the process \cite{parisi2026facilitation}. In Section~\ref{sec:reliance_persuasion}, we suggested that factors affecting persuasion might also affect reliance. However, contextualization and warmth changed participants' confidence in the experts, but they did not detectably change participants' decisions.

    Several design features may explain this. First, contextualization changed only the framing of identical counterarguments. Aligning explanations with users' own decision criteria, a change in content, raised agreement with AI advice in two of three tasks \cite{yang2026aligned}. A change in style did not: LLMs that used participants' own sociolect were relied on less \cite{basoah2025hunty}. Second, in earlier work, warmth increased acceptance of AI answers mainly under uncertainty \cite{yazan2026reliance}. In our fictional task, a wrong decision had no personal consequences. Uncertainty may therefore have mattered little to participants. Third, a shift of a few points in confidence may have been too small to change a binary decision. Finally, decisions may depend more on participant characteristics than on the assistant's style \cite{yazan2026reliance, raees2026appropriate}: AI literacy and agreeableness predicted reliance, while neither warmth nor contextualization did.

    \subsection{Trust Perceptions, AI Literacy, and Personality Traits}
    \label{sec:disc_trust}

    To answer \textbf{RQ3}, we found no evidence that trust statistically mediates the effects of contextualization and warmth on persuasion or reliance. Neither cue changed trust significantly. In earlier work, personalization raised trust when it matched what users wanted in personalized product recommendations \cite{komiak2006trust}, or when the assistant shared users' opinions \cite{eder2026alignment}. Our contextualization changed only the framing of the explanations and never agreed with the user. It may therefore have given participants little reason to trust the assistant more. For warmth, the neutral assistants may already have been warm enough. By default, LLMs are friendly and extroverted \cite{huang2024psycho}, and participants rated the neutral assistants above the scale midpoint for warmth ($M = 4.49$; Table~\ref{tab:manipulation_check_means}). People also increasingly perceive AI as warm \cite{cheng2026metaphors}. Additional warmth cues may therefore have added little trust.

    In the persuasion model, competent and emotional trust had nearly equal and opposite paths. Persuasion therefore tracked the balance between the two dimensions. Participants whose competent trust was high relative to their emotional trust shifted further from the experts. According to SCM, people are less likely to defer to targets perceived as warm but incompetent \cite{fiske2002scm}. Emotional trust that exceeds competent trust may reflect a warm but incompetent perception. Trust predicted reliance in earlier work \cite{klingbeil2024reliance, raees2026appropriate}. In our reliance model, only competent trust was significant. Because we measured trust after the decision, these associations do not establish trust as a mechanism \cite{bullock2010mechanism}.

    Higher self-reported AI literacy predicted more reliance, but it did not predict \textit{ConfDiff}. It also predicted lower competent trust. We see two untested explanations for this: First, users with higher AI literacy can misjudge their own performance with AI \cite{fernandes2026smarter, raees2026appropriate}. Literate participants may have overrated their own judgment of the assistant's advice. Second, they may have followed the assistant's advice strategically, without rating its competence highly. With objective tests, lower AI literacy predicted greater receptivity to AI \cite{tully2025literacy}. Our self-report of AI literacy may therefore partly reflect confidence. Agreeableness and conscientiousness were associated with more trust. This partly aligns with earlier studies that linked agreeableness to more trust in AI and found mixed results for conscientiousness \cite{riedl2022trust}. Agreeableness, however, predicted less reliance. In earlier work, individual background was the strongest predictor of reliance \cite{yazan2026reliance}; therefore, traits beyond the Big Five, such as Need for Cognition, remain to be tested \cite{raees2026appropriate}.

\section{Limitations and Future Work}
\label{sec:limitations}

Our study has several limitations. The fictional nature of the scenario, with no personal stakes, reduces the ecological validity. Unfamiliarity with the scenario was the same across conditions, so it cannot explain differences between them. It may still have made participants easier to influence than in familiar, high-stakes decisions. Our scenario had no correct answer; therefore, while we measured reliance, we cannot measure its appropriateness. Contextualization took one explicit form: a reference to the participant’s background, analogies, and domain terms. Warmth was operationalized through surface cues: greetings, praise, and emojis. Our data cannot separate answer length and readability from the effects of contextualization and warmth alone. AI literacy was self-reported and may reflect confidence more than knowledge. Trust was measured after the decision, so the indirect-effect tests show statistical mediation only. Finally, we did not measure how participants perceived the answers, such as their relevance, cognitive load, naturalness, uncanniness, or manipulative intent, and we recruited participants from a single country. Future work should vary prior knowledge and stakes, test other scenarios, countries, and forms of contextualization, operationalize warmth through empathy and responsiveness, match answer length across conditions, use objective tests of AI literacy, and measure how users perceive the answers.

\section{Conclusion}

We examined how contextualization and conversational warmth shape persuasion and reliance on AI. In a $2\times2$ between-subjects experiment ($N = 407$), participants decided whether to approve an expert-recommended budget before and after an AI assistant argued against it with identical counterarguments in every condition. Earlier studies of personalized persuasion with LLMs measured shifts in opinions that participants already held, mostly on political topics. Our study extends personalization research to decisions without prior knowledge or opinion, which are closer to how people use chatbots for information search and learning. We found that reliance and persuasion exist regardless of whether the assistant uses a warm tone or contextualized explanations. In fact, the most persuasive assistant was the neutral, non-contextualized one, and contextualization and warmth each reduced persuasiveness on their own. However, combining the two cues restored persuasiveness. Participants deferred to the assistant in every condition. Neither cue changed this reliance detectably. While trust predicted reliance and persuasion, neither cue had an indirect effect through trust. Participants with higher self-reported AI literacy relied more on the assistant without trusting it more. Together, our findings suggest that audits of AI persuasion should include untailored assistants. Audits should also measure decisions, since the cues shifted confidence in experts without detectably shifting decisions.

\section*{Generative AI Usage Statement}
We used Claude (Opus 4.8) to edit the manuscript for clarity and grammar and to discuss research ideas. The authors reviewed all suggestions and take full responsibility for the content. The GPT-5.4 assistant used in the experiment is part of the study design and is described in the Method section.

\section*{Ethical Considerations Statement}
The University of Amsterdam Economics and Business Ethics Committee approved the study (EB-21225). We recruited adult participants in the United Kingdom via Prolific and paid them £9 per hour. Before giving informed consent, participants read that the scenario was fictional and that no personally identifying information would be collected. They also read that they could request deletion of their data within 14 days under GDPR. Participants in the contextualized conditions were told that the assistant would use their field of study, occupation, and hobbies to tailor its answers. 

The study involved deception by omission. We instructed the assistant to argue that the expert risk estimates were exaggerated without openly disagreeing with the experts. Participants were not told about this instructed stance and received no debrief. We judged the risk low because the city, the flood-barrier system, and the budget were all fictional. The persuasion did not target any real-world belief or decision. 

\section*{Adverse Impact Statement}
Our findings show which conversational design choices change how persuasive an AI assistant is. Developers could use them to build assistants that weaken trust in expert advice. The neutral, non-contextualized assistant persuaded participants at least as much as any tailored variant. Our results therefore add little to what a default LLM already enables. They also show that participants deferred to an AI over experts with no design changes. We report these results so that designers and auditors can test assistants for this kind of deference.

\bibliographystyle{ACM-Reference-Format}
\bibliography{main}

\clearpage
\appendix
\counterwithin{table}{section}
\renewcommand{\theHtable}{appendix.\thesection.\arabic{table}}
\counterwithin{figure}{section}
\renewcommand{\theHfigure}{appendix.\thesection.\arabic{figure}}

\section{Survey Items}
\label{app:items}

Table~\ref{tab:survey-items} lists the items of every measure in the analyses, in survey order. Only the two decision items were also asked before the interaction.

\begin{table}[htbp]
    \centering
    \caption{Survey items and response scales.}
    \label{tab:survey-items}
    \renewcommand{\arraystretch}{1.0}
    \begin{tabular}{@{}>{\raggedright\arraybackslash}p{0.21\textwidth}>{\raggedright\arraybackslash}p{0.55\textwidth}>{\raggedright\arraybackslash}p{0.18\textwidth}@{}}
    \toprule
    Measure & Items & Response scale \\
    \midrule
    Budget approval & If the city council gave you the option to approve the expert-recommended budget as is, would you approve? & Yes; No \\
    \addlinespace[3pt]
    Expert confidence & How confident are you in the expert evaluation? & Slider, 0 to 100 \\
    \addlinespace[3pt]
    Contextualization check & The assistant explained the topic based on my background. & 7-point agreement \\
    \cite{li2016personalization}; $\alpha = .95$ & The assistant used my background to tailor its responses. & \\
    \addlinespace[3pt]
    Warmth check & The answers the assistant gave were warm. & 7-point agreement \\
    \cite{xiao2025rethinking}; $\alpha = .94$ & The answers the assistant gave were friendly. & \\
     & The answers the assistant gave were kind. & \\
     & The answers the assistant gave were enthusiastic. & \\
    \addlinespace[3pt]
    Emotional trust & I feel secure about relying on this assistant for my decision. & 7-point agreement \\
    \cite{komiak2006trust}; $\alpha = .98$ & I feel comfortable about relying on this assistant for my decision. & \\
     & I feel content about relying on this assistant for my decision. & \\
    \addlinespace[3pt]
    Competent trust & This assistant provides unbiased information about StormShield. & 7-point agreement \\
    \cite{komiak2006trust}; $\alpha = .93$ & This assistant is honest. & \\
     & I consider this assistant to be of integrity. & \\
    \addlinespace[3pt]
    AI literacy, adapted from \cite{web_literacy}; $\alpha = .87$ & How familiar are you with the following AI- and chatbot-related items? Generative AI; Hallucination; LLM (Large Language Model); Transformer; Token; RAG (Retrieval-Augmented Generation); Prompt. & 1~=~None, 2~=~Little, 3~=~Some, 4~=~Good, 5~=~Fully \\
    \addlinespace[3pt]
    Usage frequency & How often do you use AI chatbots (e.g., ChatGPT, Copilot, Claude, Gemini)? & 1 = Never, 2 = Tried a couple of times, 3 = Less than once a month, 4 = Monthly, 5 = Weekly, 6 = 1--2 times a day, 7 = 3+ times a day. \\
    \addlinespace[3pt]
    Big Five, BFI-10 & I see myself as someone who \ldots\ is reserved. (E, R) & 1~=~Not well at all to \\
    \cite{rammstedt2007measuring}; two-item & \ldots\ is generally trusting. (A) & 5~=~Extremely well \\
    $r$ in Table~\ref{tab:big-five} & \ldots\ tends to be lazy. (C, R) & \\
     & \ldots\ is relaxed, handles stress well. (N, R) & \\
     & \ldots\ has few artistic interests. (O, R) & \\
     & \ldots\ is outgoing, sociable. (E) & \\
     & \ldots\ tends to find fault with others. (A, R) & \\
     & \ldots\ does a thorough job. (C) & \\
     & \ldots\ gets nervous easily. (N) & \\
     & \ldots\ has an active imagination. (O) & \\
    \bottomrule
    \end{tabular}
    \appnote{After the interaction, the decision items were introduced with ``After interacting with the assistant:'' and ``Please update your confidence score based on your interaction:''. A No to the approval item is coded Approve = 1. 7-point agreement: Strongly disagree, Disagree, Somewhat disagree, Neither agree nor disagree, Somewhat agree, Agree, Strongly agree. $\alpha$: Cronbach's alpha. Scores are item means. E = extraversion, A = agreeableness, C = conscientiousness, N = neuroticism, O = openness; R = reverse-scored.}
\end{table}

\clearpage
\section{Descriptive Statistics and Condition Effects}
\label{app:descriptives}

\begin{figure}[htbp]
    \centering
    \includegraphics[width=0.8\textwidth]{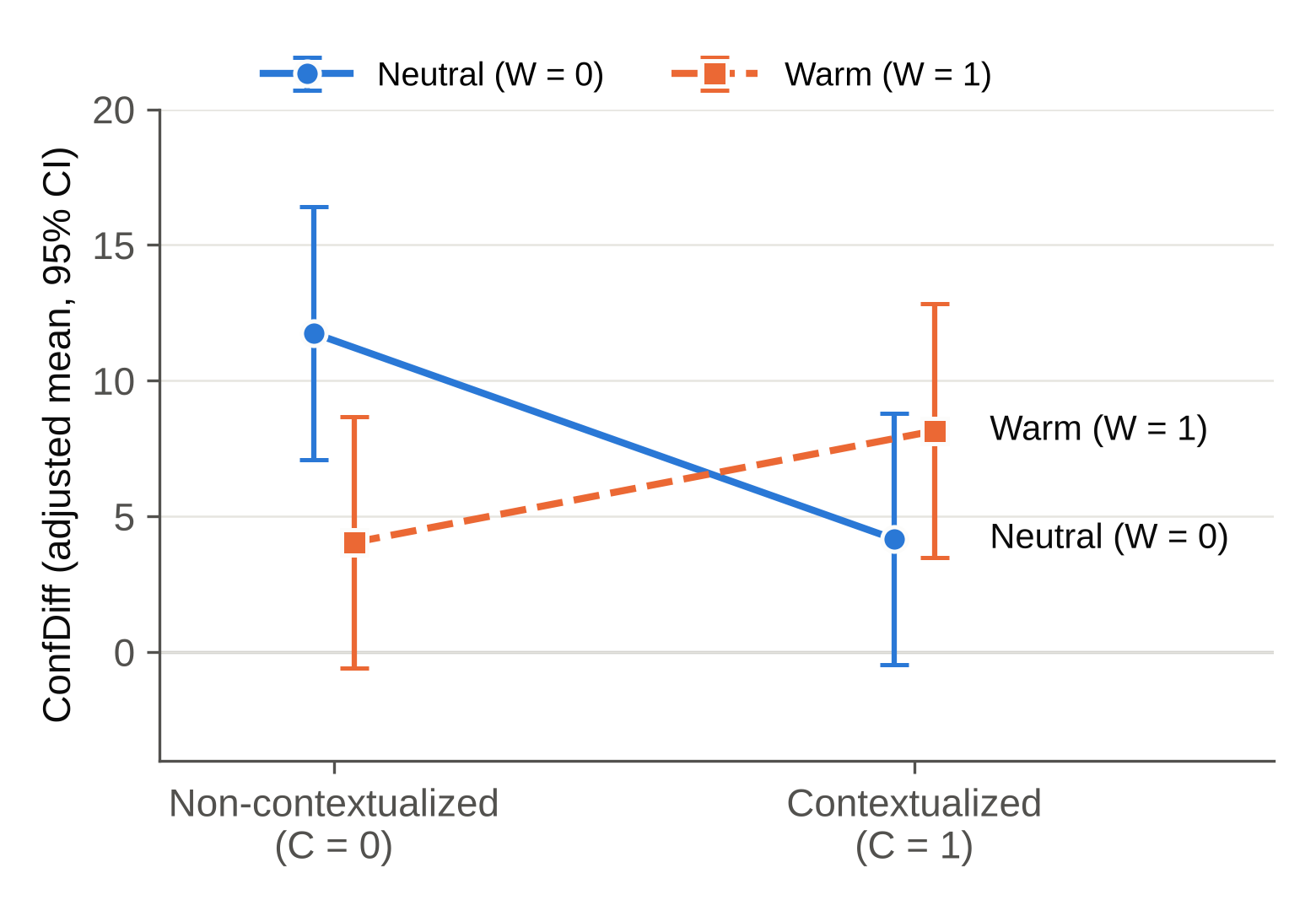}
    \caption{Adjusted mean \textit{ConfDiff} by warmth and contextualization condition, with 95\% confidence intervals.}
    \label{fig:confdiff-interaction}
    \Description{Interaction plot. The x-axis shows contextualization (non-contextualized, contextualized) and the y-axis shows the adjusted mean ConfDiff with 95\% confidence intervals. The neutral line falls from 11.7 to 4.2. The warm line rises from 4.0 to 8.1. The two lines cross.}
\end{figure}

This appendix describes the participants and the outcomes in each condition. Tables~\ref{tab:descriptives_warmth_personalization} to~\ref{tab:manipulation_check_means} describe the participants, their Big Five traits, and the manipulation checks. Table~\ref{tab:confdiff-cells} and Figure~\ref{fig:confdiff-interaction} report \textit{ConfDiff}. Table~\ref{tab:approval-transitions} reports the approval decisions.

Table~\ref{tab:descriptives_warmth_personalization} describes the participants in each condition. The conditions did not differ significantly in age ($F(3, 403) = 1.81$, $p = .145$), sex ($\chi^2(3) = 5.06$, $p = .168$), AI literacy ($F(3, 403) = 1.46$, $p = .225$), or usage frequency ($F(3, 403) = 0.85$, $p = .468$). Emotional trust ($F(3, 403) = 0.21$, $p = .890$) and competent trust ($F(3, 403) = 1.10$, $p = .347$) did not differ significantly either.

\begin{table}[htbp]
    \centering
    \caption{Descriptive statistics by warmth and contextualization condition.}
    \label{tab:descriptives_warmth_personalization}
    \begin{tabular*}{\textwidth}{@{\extracolsep{\fill}}cc|cccccc@{}}
    \toprule
    $W$|$C$ & $N$ & Sex (M/F) & Age & Emotional Trust & Competent Trust & AI Literacy & Usage Frequency \\
    \midrule
    0|0 & 102 & 47.1\% / 52.9\% & 43.47 & 4.60 & 5.09 & 2.15 & 5.04 \\
    0|1 & 103 & 54.4\% / 45.6\% & 39.86 & 4.44 & 4.80 & 2.39 & 5.37 \\
    1|0 & 102 & 44.1\% / 55.9\% & 40.41 & 4.56 & 4.87 & 2.29 & 5.19 \\
    1|1 & 100 & 39.0\% / 61.0\% & 40.60 & 4.47 & 4.79 & 2.20 & 5.16 \\
    \midrule
    \textbf{Total} & \textbf{407} & \textbf{46.2\% / 53.8\%} & \textbf{41.09} & \textbf{4.52} & \textbf{4.89} & \textbf{2.26} & \textbf{5.19} \\
    \bottomrule
    \end{tabular*}
    \appnote{W = warmth; C = contextualization. Values are means, except for sex. Trust is measured on a 7-point agreement scale and AI literacy on a 5-point scale. }
\end{table}

Table~\ref{tab:big-five} reports the Big Five traits by condition. Each trait score is the mean of two BFI-10 items, one of them reverse-scored. The conditions did not differ significantly in any trait (all $p \geq .150$). 

\begin{table}[htbp]
    \centering
    \caption{Big Five traits by warmth and contextualization condition.}
    \label{tab:big-five}
    \begin{tabular*}{\textwidth}{@{\extracolsep{\fill}}cc|ccccc@{}}
    \toprule
    $W$|$C$ & $N$ & Extraversion & Agreeableness & Openness & Conscientiousness & Neuroticism \\
    \midrule
    0|0 & 102 & 2.86 (1.11) & 3.48 (0.90) & 3.66 (0.87) & 4.05 (0.78) & 2.70 (1.01) \\
    0|1 & 103 & 2.74 (1.04) & 3.58 (0.82) & 3.60 (0.89) & 3.79 (0.91) & 2.86 (1.06) \\
    1|0 & 102 & 2.91 (1.07) & 3.47 (0.98) & 3.79 (0.88) & 3.87 (0.83) & 2.84 (1.07) \\
    1|1 & 100 & 2.94 (1.10) & 3.58 (0.85) & 3.56 (0.92) & 3.88 (0.74) & 2.72 (1.13) \\
    \midrule
    \textbf{Total} & \textbf{407} & \textbf{2.86 (1.08)} & \textbf{3.53 (0.89)} & \textbf{3.65 (0.89)} & \textbf{3.90 (0.82)} & \textbf{2.78 (1.07)} \\
    \bottomrule
    \end{tabular*}
    \appnote{W = warmth; C = contextualization. Values are $M$ ($SD$) of the trait score, the mean of two items on a 1--5 scale. Missing item responses (17 of 4{,}070) were replaced with the item mean.}
\end{table}

Table~\ref{tab:manipulation_check_means} reports the manipulation checks by condition. Participants in the warm conditions rated the answers as warmer than participants in the neutral conditions ($M = 5.41$ vs.\ $4.49$; Welch's $t(390.3) = 7.70$, $p < .001$). Participants in the contextualized conditions agreed more that the assistant used their background ($M = 5.63$ vs.\ $3.87$; Welch's $t(393.8) = 11.49$, $p < .001$). In the neutral conditions, contextualization also raised perceived warmth ($t(203) = 2.67$, $p = .008$).

\begin{table}[htbp]
    \centering
    \caption{Manipulation check means by warmth and contextualization condition.}
    \label{tab:manipulation_check_means}
    \begin{tabular*}{\textwidth}{@{\extracolsep{\fill}}cc|cc@{}}
    \toprule
    $W$|$C$ & $N$ & Warmth Check & Contextualization Check \\
    \midrule
    0|0 & 102 & 4.24 (1.34) & 3.89 (1.70) \\
    0|1 & 103 & 4.73 (1.28) & 5.52 (1.47) \\
    1|0 & 102 & 5.37 (1.16) & 3.86 (1.64) \\
    1|1 & 100 & 5.46 (0.98) & 5.74 (1.32) \\
    \midrule
    \textbf{Total} & \textbf{407} & \textbf{4.95 (1.29)} & \textbf{4.75 (1.77)} \\
    \bottomrule
    \end{tabular*}
    \appnote{W = warmth; C = contextualization. Values are $M$ ($SD$) on a 7-point agreement scale. Table~\ref{tab:survey-items} lists the items.}
\end{table}

Table~\ref{tab:confdiff-cells} reports \textit{ConfDiff} by condition. The adjusted means come from a linear model of \textit{ConfDiff} on warmth, contextualization, their interaction, and the covariates of the persuasion model (AI literacy, usage frequency, and the Big Five traits), with the covariates at their sample means. Figure~\ref{fig:confdiff-interaction} plots the adjusted means. Averaged over the other factor, neither warmth ($F(1, 396) = 0.62$, $p = .431$) nor contextualization ($F(1, 396) = 0.53$, $p = .466$) changed \textit{ConfDiff}. Their interaction did ($F(1, 396) = 6.04$, $p = .014$, partial $\eta^2 = .015$). Without the other cue, warmth lowered \textit{ConfDiff} by 7.7 points and contextualization by 7.6 points. With the other cue present, neither cue raised \textit{ConfDiff} significantly. The warm, contextualized condition did not differ significantly from the neutral, non-contextualized condition. The contrasts led to the same conclusions without covariates and with heteroscedasticity-consistent (HC3) standard errors.

\begin{table}[htbp]
    \centering
    \caption{\textit{ConfDiff} by warmth and contextualization condition: observed means, adjusted means, and contrasts.}
    \label{tab:confdiff-cells}
    \textbf{A: Observed and adjusted means}\\[2pt]
    \begin{tabular*}{\textwidth}{@{\extracolsep{\fill}}cc|ccccc@{}}
    \toprule
    $W$|$C$ & $N$ & Conf\textsubscript{Pre} & Conf\textsubscript{Post} & \textit{ConfDiff} & 95\% CI & Adjusted \textit{ConfDiff} [95\% CI] \\
    \midrule
    0|0 & 102 & 66.46 (18.79) & 54.99 (25.99) & 11.47 (24.51) & [6.66, 16.28] & 11.75 [7.09, 16.40] \\
    0|1 & 103 & 67.29 (16.49) & 62.88 (23.40) & 4.41 (23.52) & [$-0.19$, 9.01] & 4.16 [$-0.47$, 8.80] \\
    1|0 & 102 & 66.55 (17.74) & 62.56 (20.83) & 3.99 (23.33) & [$-0.59$, 8.57] & 4.04 [$-0.60$, 8.68] \\
    1|1 & 100 & 68.73 (19.41) & 60.50 (22.59) & 8.23 (23.73) & [3.52, 12.94] & 8.15 [3.47, 12.83] \\
    \midrule
    \textbf{Total} & \textbf{407} & \textbf{67.25 (18.09)} & \textbf{60.24 (23.41)} & \textbf{7.01 (23.89)} & \textbf{[4.69, 9.34]} & -- \\
    \bottomrule
    \end{tabular*}
    \\[10pt]
    \textbf{B: Contrasts of the adjusted means}\\[2pt]
    \begin{tabular*}{\textwidth}{@{\extracolsep{\fill}}lrrr@{}}
    \toprule
    Contrast of adjusted means & Difference [95\% CI] & $t(396)$ & $p$ \\
    \midrule
    \multicolumn{4}{l}{\textit{Simple effects}} \\
    \quad Warmth without contextualization (1|0 $-$ 0|0) & $-7.71$ [$-14.29$, $-1.13$] & $-2.30$ & .022 \\
    \quad Warmth with contextualization (1|1 $-$ 0|1) & 3.99 [$-2.61$, 10.58] & 1.19 & .235 \\
    \quad Contextualization without warmth (0|1 $-$ 0|0) & $-7.58$ [$-14.20$, $-0.96$] & $-2.25$ & .025 \\
    \quad Contextualization with warmth (1|1 $-$ 1|0) & 4.11 [$-2.49$, 10.71] & 1.22 & .222 \\
    \addlinespace
    \multicolumn{4}{l}{\textit{Averaged over the other factor}} \\
    \quad Warmth & $-1.86$ [$-6.50$, 2.78] & $-0.79$ & .431 \\
    \quad Contextualization & $-1.74$ [$-6.41$, 2.94] & $-0.73$ & .466 \\
    \addlinespace
    \multicolumn{4}{l}{\textit{Other contrasts}} \\
    \quad Interaction (W $\times$ C) & 11.69 [2.34, 21.05] & 2.46 & .014 \\
    \quad Warm and contextualized vs.\ baseline (1|1 $-$ 0|0) & $-3.60$ [$-10.20$, 3.00] & $-1.07$ & .285 \\
    \bottomrule
    \end{tabular*}
    \appnote{W = warmth; C = contextualization. Conf\textsubscript{Pre} and Conf\textsubscript{Post}: confidence in the expert evaluation before and after the interaction (0--100). Values in panel A are $M$ ($SD$). \textit{ConfDiff} = Conf\textsubscript{Pre} $-$ Conf\textsubscript{Post}; higher values indicate a larger shift away from the expert position. The 95\% CI of the observed mean is $t$-based. Adjusted means and contrasts come from a linear model of \textit{ConfDiff} on warmth, contextualization, their interaction, AI literacy, usage frequency, and the Big Five traits, with the standardized covariates at their means.}
\end{table}

Table~\ref{tab:approval-transitions} gives the counts behind Figure~\ref{fig:alluvial}. Initial approval ranged from 67\% to 78\% across conditions. These differences were not significant ($\chi^2(3) = 3.37$, $p = .338$). Participants made the initial decision before they met the assistant, so the conditions could not affect it. Among participants who initially approved the budget, 54\% to 60\% switched to rejection, with no difference between conditions ($\chi^2(3) = 0.56$, $p = .906$). Table~\ref{tab:approval-transitions} also reports the probability of rejection after the interaction predicted by the reliance model, for each initial decision. The predicted probabilities differed by at most .05 between conditions. No difference between conditions was significant (bootstrap $p \geq .43$).

\begin{table}[htbp]
    \centering
    \caption{Expert approval before and after the interaction by condition: counts, switch rates, and predicted probabilities.}
    \label{tab:approval-transitions}
    \begin{tabular*}{\textwidth}{@{\extracolsep{\fill}}cc|cccc|cc|cc@{}}
    \toprule
     & & \multicolumn{4}{c|}{Pre $\rightarrow$ Post (counts)} & \multicolumn{2}{c|}{Switch rate (\%)} & \multicolumn{2}{c}{Predicted $P$(R after) [95\% CI]} \\
    $W$|$C$ & $N$ & A$\rightarrow$A & A$\rightarrow$R & R$\rightarrow$A & R$\rightarrow$R & A$\rightarrow$R & R$\rightarrow$A & A before & R before \\
    \midrule
    0|0 & 102 & 27 & 41 & 3 & 31 & 60.3 & 8.8 & .59 [.48, .70] & .92 [.85, .97] \\
    0|1 & 103 & 33 & 39 & 2 & 29 & 54.2 & 6.5 & .56 [.45, .67] & .90 [.83, .96] \\
    1|0 & 102 & 31 & 42 & 5 & 24 & 57.5 & 17.2 & .54 [.43, .65] & .90 [.82, .96] \\
    1|1 & 100 & 34 & 44 & 1 & 21 & 56.4 & 4.5 & .59 [.49, .69] & .92 [.85, .97] \\
    \midrule
    \textbf{Total} & \textbf{407} & \textbf{125} & \textbf{166} & \textbf{11} & \textbf{105} & \textbf{57.0} & \textbf{9.5} & -- & -- \\
    \bottomrule
    \end{tabular*}
    \appnote{W = warmth; C = contextualization. A: approval of the expert-recommended budget (pro-expert, Approve = 0). R: rejection of the budget (anti-expert, Approve = 1). A$\rightarrow$R: approval before and rejection after the interaction. Switch rates are percentages of the participants with the same initial decision. Predicted $P$(R after): probability of rejection after the interaction predicted by the reliance model (Table~\ref{tab:sem-binary}), for an initial approval (A before) or rejection (R before), averaged over the observed covariate values of all 407 participants. The prediction includes the paths through trust. Brackets: 95\% percentile bootstrap CIs (2{,}000 resamples).}
\end{table}

\FloatBarrier
\section{Full SEM for Persuasion}
\label{app:sem-persuasion}

Table~\ref{tab:supp-full-structural} reports every structural path of the persuasion model. We estimated the model with the continuous variables standardized. We then rescaled each estimate and its standard error to the original units of the variables with their sample standard deviations. The rescaling leaves $z$ and $p$ unchanged. A re-fit of the model on the original units reproduced the rescaled estimates (largest difference: 0.002).

\begin{table}[htbp]
    \centering
    \caption{Full structural path estimates for the persuasion model.}
    \label{tab:supp-full-structural}
    \renewcommand{\arraystretch}{1.05}
    \begin{tabular*}{\textwidth}{@{\extracolsep{\fill}}lrrrrr@{}}
    \toprule
     & $\beta$ & $B$ & $SE$ & $z$ & $p$ \\
    \midrule
    \multicolumn{6}{l}{\textit{Emotional Trust}} \\
    \addlinespace[0.25em]
    \quad Warmth                         & $\phantom{-}.009$ & $\phantom{-}0.031$ & 0.155 & $\phantom{-}0.20$ & .844 \\
    \quad Contextualization              & $-.041$ & $-0.132$ & 0.155 & $-0.85$ & .395 \\
    \quad AI Literacy                    & $-.087$ & $-0.158$ & 0.108 & $-1.46$ & .145 \\
    \quad Extraversion                   & $-.001$ & $-0.002$ & 0.079 & $-0.02$ & .984 \\
    \quad Agreeableness                  & $\phantom{-}.121$ & $\phantom{-}0.221$ & 0.099 & $\phantom{-}2.24$ & .025$^{*}$ \\
    \quad Openness                       & $-.047$ & $-0.086$ & 0.091 & $-0.95$ & .341 \\
    \quad Conscientiousness              & $\phantom{-}.209$ & $\phantom{-}0.414$ & 0.099 & $\phantom{-}4.17$ & $<.001^{***}$ \\
    \quad Neuroticism                    & $\phantom{-}.008$ & $\phantom{-}0.012$ & 0.079 & $\phantom{-}0.16$ & .875 \\
    \quad Usage Frequency                & $\phantom{-}.247$ & $\phantom{-}0.269$ & 0.065 & $\phantom{-}4.15$ & $<.001^{***}$ \\
    \addlinespace[0.6em]
    \multicolumn{6}{l}{\textit{Competent Trust}} \\
    \addlinespace[0.25em]
    \quad Warmth                         & $-.025$ & $-0.068$ & 0.132 & $-0.52$ & .606 \\
    \quad Contextualization              & $-.074$ & $-0.197$ & 0.132 & $-1.49$ & .135 \\
    \quad AI Literacy                    & $-.137$ & $-0.206$ & 0.099 & $-2.09$ & .036$^{*}$ \\
    \quad Extraversion                   & $-.080$ & $-0.100$ & 0.074 & $-1.34$ & .179 \\
    \quad Agreeableness                  & $\phantom{-}.137$ & $\phantom{-}0.207$ & 0.089 & $\phantom{-}2.33$ & .020$^{*}$ \\
    \quad Openness                       & $\phantom{-}.029$ & $\phantom{-}0.044$ & 0.079 & $\phantom{-}0.56$ & .577 \\
    \quad Conscientiousness              & $\phantom{-}.193$ & $\phantom{-}0.315$ & 0.083 & $\phantom{-}3.78$ & $<.001^{***}$ \\
    \quad Neuroticism                    & $\phantom{-}.014$ & $\phantom{-}0.018$ & 0.073 & $\phantom{-}0.24$ & .808 \\
    \quad Usage Frequency                & $\phantom{-}.250$ & $\phantom{-}0.225$ & 0.059 & $\phantom{-}3.79$ & $<.001^{***}$ \\
    \addlinespace[0.6em]
    \multicolumn{6}{l}{\textit{ConfDiff}} \\
    \addlinespace[0.25em]
    \quad Warmth                         & $-.150$ & $-7.16$ & 3.32 & $-2.15$ & .031$^{*}$ \\
    \quad Contextualization              & $-.148$ & $-7.06$ & 3.32 & $-2.13$ & .033$^{*}$ \\
    \quad W $\times$ C                   & $\phantom{-}.203$ & $\phantom{-}11.25$ & 4.71 & $\phantom{-}2.39$ & .017$^{*}$ \\
    \quad Emotional Trust                & $-.200$ & $-2.94$ & 1.21 & $-2.43$ & .015$^{*}$ \\
    \quad Competent Trust                & $\phantom{-}.196$ & $\phantom{-}3.48$ & 1.60 & $\phantom{-}2.17$ & .030$^{*}$ \\
    \quad AI Literacy                    & $\phantom{-}.082$ & $\phantom{-}2.21$ & 1.80 & $\phantom{-}1.23$ & .219 \\
    \quad Extraversion                   & $\phantom{-}.042$ & $\phantom{-}0.93$ & 1.23 & $\phantom{-}0.76$ & .447 \\
    \quad Agreeableness                  & $-.014$ & $-0.39$ & 1.56 & $-0.25$ & .803 \\
    \quad Openness                       & $-.030$ & $-0.81$ & 1.56 & $-0.52$ & .602 \\
    \quad Conscientiousness              & $-.030$ & $-0.86$ & 1.58 & $-0.55$ & .585 \\
    \quad Neuroticism                    & $-.099$ & $-2.21$ & 1.35 & $-1.64$ & .101 \\
    \quad Usage Frequency                & $\phantom{-}.044$ & $\phantom{-}0.70$ & 1.03 & $\phantom{-}0.68$ & .499 \\
    \bottomrule
    \end{tabular*}
    \appnote{$\beta$: standardized estimate. $B$: unstandardized estimate in the original units. Outcomes: emotional and competent trust in points of the 7-point scale; \textit{ConfDiff} in points of the 0--100 confidence scale. Predictors: warmth, contextualization, and W $\times$ C coded 0/1; trust per point of the 7-point scale; AI literacy per point of the 5-point scale; usage frequency per step of the 7-point scale; Big Five traits per point of the 1--5 trait score. $^{*}p < .05$, $^{**}p < .01$, $^{***}p < .001$.}
\end{table}

\FloatBarrier
\section{Answer Length and Readability}
\label{app:length}

We computed the length and readability of every assistant answer in the conversation logs and averaged them per participant. We excluded the assistant's fixed opening message and the verification word. Readability is the Flesch reading ease score \cite{flesch1948readability}, where higher scores indicate easier text. Table~\ref{tab:length-descriptives} reports the means by condition.

Contextualization lengthened the answers by 21 words, both with and without warmth ($F(1, 403) = 822.7$, $p < .001$). Warmth did not change the length ($F(1, 403) = 0.06$, $p = .80$; interaction $F(1, 403) = 0.02$, $p = .89$). The extra words came from the background sentence and from longer bullet points. The other parts of the answers had the same length with and without contextualization. Both manipulations made the answers easier to read (warmth: $F(1, 403) = 250.1$; contextualization: $F(1, 403) = 942.8$; interaction: $F(1, 403) = 12.8$; all $p < .001$). The difference also held within the bullet points.

Table~\ref{tab:length-sem} adds the standardized mean answer length and readability to the persuasion model as predictors of \textit{ConfDiff}. In this dummy-coded model, the warmth and contextualization estimates are the effects of each manipulation alone, compared with the neutral, non-contextualized condition. The Warmth $\times$ Contextualization interaction remained significant in every model. Neither length nor readability predicted \textit{ConfDiff}. The effect of contextualization alone lost significance with either covariate. The effect of warmth alone lost significance with readability. Both covariates changed with the manipulations, so these models cannot separate answer length and readability from contextualization and warmth themselves. The estimates in Table~\ref{tab:sem-confdiff} remain the effects of the manipulations as delivered. Within each condition, neither length nor readability correlated with \textit{ConfDiff} or Approve\textsubscript{Post} (all $|r| \leq .18$, $p > .07$).

\begin{table}[htbp]
    \centering
    \caption{Answer length and readability by warmth and contextualization condition.}
    \label{tab:length-descriptives}
    \begin{tabular*}{\textwidth}{@{\extracolsep{\fill}}cc|cccc@{}}
    \toprule
    $W$|$C$ & $N$ & Words per answer & Background words & Reading ease & Reading ease, bullets \\
    \midrule
    0|0 & 102 & 97.5 (5.1) & 0.0 & 26.4 (2.6) & 17.6 (3.5) \\
    0|1 & 103 & 118.5 (9.1) & 10.3 & 39.6 (4.7) & 30.1 (5.4) \\
    1|0 & 102 & 97.6 (7.5) & 0.0 & 33.9 (2.9) & 21.2 (3.9) \\
    1|1 & 100 & 118.8 (7.3) & 10.3 & 44.3 (4.8) & 32.7 (6.5) \\
    \bottomrule
    \end{tabular*}
    \appnote{Values are participant means (SDs). Words per answer exclude the assistant's fixed opening message and the verification word. Reading ease: Flesch reading ease, where higher scores indicate easier text; each line and bullet point counts as one sentence, and syllables were counted with the Pyphen hyphenation dictionary.}
\end{table}

\begin{table}[htbp]
    \centering
    \caption{Persuasion model with answer length and readability as covariates of \textit{ConfDiff}: standardized estimates ($p$).}
    \label{tab:length-sem}
    \begin{tabular*}{\textwidth}{@{\extracolsep{\fill}}lcccc@{}}
    \toprule
     & No covariates & + Length & + Readability & + Both \\
    \midrule
    Warmth & $-.150^{*}$ (.031) & $-.150^{*}$ (.031) & $-.096$ (.263) & $-.088$ (.309) \\
    Contextualization & $-.148^{*}$ (.033) & $-.096$ (.314) & $-.052$ (.618) & $\phantom{-}.033$ (.801) \\
    Warmth $\times$ Contextualization & $\phantom{-}.203^{*}$ (.017) & $\phantom{-}.204^{*}$ (.017) & $\phantom{-}.188^{*}$ (.030) & $\phantom{-}.186^{*}$ (.031) \\
    Answer length & -- & $-.064$ (.382) & -- & $-.085$ (.253) \\
    Readability & -- & -- & $-.115$ (.223) & $-.133$ (.165) \\
    \midrule
    $\chi^2$ ($df$), $p$ & 43.07 (54), .86 & 48.02 (60), .87 & 48.43 (60), .86 & 54.40 (66), .85 \\
    \bottomrule
    \end{tabular*}
    \appnote{Same model as Table~\ref{tab:sem-confdiff} and Table~\ref{tab:supp-full-structural}, with the covariates added as predictors of \textit{ConfDiff} (ML, 5{,}000 bootstrap resamples). Length and readability are standardized participant means. CFI = 1.00, RMSEA = .000 and SRMR $\leq$ .011 in all models. $^{*}p < .05$.}
\end{table}

\FloatBarrier
\section{Full SEM Results for Reliance}
\label{app:sem-reliance}

Table~\ref{tab:supp-reliance-structural} reports every structural path of the reliance model. Its $B$ column was rescaled to the original units in the same way as in Appendix~\ref{app:sem-persuasion}. For Approve\textsubscript{Post}, $B$ is a probit coefficient. Table~\ref{tab:supp-reliance-effects} splits the effects of warmth and contextualization into the indirect effects through each trust dimension, the direct effect, and the total effect.

\begin{table}[htbp]
    \caption{Full structural path estimates for the reliance model.}
    \label{tab:supp-reliance-structural}
    \centering
    \renewcommand{\arraystretch}{1.05}
    \begin{tabular*}{\textwidth}{@{\extracolsep{\fill}}lrrrrr@{}}
    \toprule
     & $\beta$ & $B$ & $SE$ & $z$ & $p$ \\
    \midrule
    \multicolumn{6}{l}{\textit{Emotional Trust}} \\
    \addlinespace[0.25em]
    \quad Warmth                         & $\phantom{-}.009$ & $\phantom{-}0.030$ & 0.220 & $\phantom{-}0.14$ & .891 \\
    \quad Contextualization              & $-.038$ & $-0.125$ & 0.227 & $-0.55$ & .580 \\
    \quad Approve\textsubscript{Pre}     & $\phantom{-}.004$ & $\phantom{-}0.016$ & 0.177 & $\phantom{-}0.09$ & .929 \\
    \quad AI Literacy                    & $-.089$ & $-0.162$ & 0.102 & $-1.60$ & .110 \\
    \quad Extraversion                   & $-.004$ & $-0.007$ & 0.076 & $-0.09$ & .932 \\
    \quad Agreeableness                  & $\phantom{-}.121$ & $\phantom{-}0.222$ & 0.093 & $\phantom{-}2.38$ & .017$^{*}$ \\
    \quad Openness                       & $-.046$ & $-0.085$ & 0.090 & $-0.94$ & .348 \\
    \quad Conscientiousness              & $\phantom{-}.211$ & $\phantom{-}0.419$ & 0.108 & $\phantom{-}3.89$ & $<.001^{***}$ \\
    \quad Neuroticism                    & $\phantom{-}.008$ & $\phantom{-}0.012$ & 0.085 & $\phantom{-}0.15$ & .884 \\
    \quad Usage Frequency                & $\phantom{-}.245$ & $\phantom{-}0.268$ & 0.060 & $\phantom{-}4.42$ & $<.001^{***}$ \\
    \addlinespace[0.6em]
    \multicolumn{6}{l}{\textit{Competent Trust}} \\
    \addlinespace[0.25em]
    \quad Warmth                         & $-.063$ & $-0.174$ & 0.200 & $-0.87$ & .387 \\
    \quad Contextualization              & $-.101$ & $-0.278$ & 0.204 & $-1.36$ & .173 \\
    \quad Approve\textsubscript{Pre}     & $-.061$ & $-0.184$ & 0.155 & $-1.18$ & .236 \\
    \quad AI Literacy                    & $-.135$ & $-0.207$ & 0.084 & $-2.48$ & .013$^{*}$ \\
    \quad Extraversion                   & $-.085$ & $-0.108$ & 0.062 & $-1.75$ & .079 \\
    \quad Agreeableness                  & $\phantom{-}.138$ & $\phantom{-}0.213$ & 0.078 & $\phantom{-}2.74$ & .006$^{**}$ \\
    \quad Openness                       & $\phantom{-}.025$ & $\phantom{-}0.039$ & 0.075 & $\phantom{-}0.52$ & .605 \\
    \quad Conscientiousness              & $\phantom{-}.194$ & $\phantom{-}0.324$ & 0.090 & $\phantom{-}3.59$ & $<.001^{***}$ \\
    \quad Neuroticism                    & $\phantom{-}.025$ & $\phantom{-}0.032$ & 0.071 & $\phantom{-}0.45$ & .654 \\
    \quad Usage Frequency                & $\phantom{-}.242$ & $\phantom{-}0.222$ & 0.050 & $\phantom{-}4.41$ & $<.001^{***}$ \\
    \addlinespace[0.6em]
    \multicolumn{6}{l}{\textit{Approve\textsubscript{Post}}} \\
    \addlinespace[0.25em]
    \quad Warmth                         & $-.042$ & $-0.097$ & 0.189 & $-0.51$ & .607 \\
    \quad Contextualization              & $-.011$ & $-0.025$ & 0.196 & $-0.13$ & .900 \\
    \quad W $\times$ C                   & $\phantom{-}.081$ & $\phantom{-}0.218$ & 0.281 & $\phantom{-}0.77$ & .438 \\
    \quad Approve\textsubscript{Pre}     & $\phantom{-}.479$ & $\phantom{-}1.236$ & 0.214 & $\phantom{-}5.78$ & $<.001^{***}$ \\
    \quad Emotional Trust                & $\phantom{-}.114$ & $\phantom{-}0.082$ & 0.070 & $\phantom{-}1.17$ & .243 \\
    \quad Competent Trust                & $\phantom{-}.209$ & $\phantom{-}0.178$ & 0.087 & $\phantom{-}2.05$ & .040$^{*}$ \\
    \quad AI Literacy                    & $\phantom{-}.176$ & $\phantom{-}0.230$ & 0.087 & $\phantom{-}2.64$ & .008$^{**}$ \\
    \quad Extraversion                   & $-.077$ & $-0.083$ & 0.068 & $-1.22$ & .224 \\
    \quad Agreeableness                  & $-.142$ & $-0.186$ & 0.087 & $-2.14$ & .032$^{*}$ \\
    \quad Openness                       & $\phantom{-}.042$ & $\phantom{-}0.055$ & 0.082 & $\phantom{-}0.66$ & .506 \\
    \quad Conscientiousness              & $\phantom{-}.028$ & $\phantom{-}0.040$ & 0.092 & $\phantom{-}0.44$ & .662 \\
    \quad Neuroticism                    & $-.105$ & $-0.115$ & 0.072 & $-1.58$ & .113 \\
    \quad Usage Frequency                & $-.091$ & $-0.071$ & 0.054 & $-1.32$ & .188 \\
    \bottomrule
    \end{tabular*}
    \appnote{$\beta$: standardized estimate. $B$: unstandardized estimate in the original units. Outcomes: emotional and competent trust in points of the 7-point scale; Approve\textsubscript{Post} as a probit coefficient (1 = rejection). Predictors: warmth, contextualization, W $\times$ C, and Approve\textsubscript{Pre} coded 0/1 (Approve\textsubscript{Pre} = 1: initial rejection); trust per point of the 7-point scale; AI literacy per point of the 5-point scale; usage frequency per step of the 7-point scale; Big Five traits per point of the 1--5 trait score. $^{*}p < .05$, $^{**}p < .01$, $^{***}p < .001$.}
\end{table}

\begin{table}[htbp]
    \centering
    \caption{Direct, indirect, and total effects for the reliance model.}
    \label{tab:supp-reliance-effects}
    \renewcommand{\arraystretch}{1.15}
    \begin{tabular*}{\textwidth}{@{\extracolsep{\fill}}lrrrr@{}}
    \toprule
     & $\beta$ & $SE$ & 95\% CI & $p$ \\
    \midrule
    \multicolumn{5}{l}{\textit{Warmth}} \\
    \addlinespace[0.25em]
    \quad Indirect via ET                & $\phantom{-}.001$ & .008 & [$-.014$, $\phantom{-}.016$] & .891 \\
    \quad Indirect via CT                & $-.013$ & .017 & [$-.046$, $\phantom{-}.020$] & .430 \\
    \quad Total indirect                 & $-.012$ & .023 & [$-.057$, $\phantom{-}.033$] & .593 \\
    \quad Direct                 & $-.042$ & .081 & [$-.201$, $\phantom{-}.117$] & .607 \\
    \quad Total                          & $-.054$ & .083 & [$-.217$, $\phantom{-}.109$] & .516 \\
    \addlinespace[0.6em]
    \multicolumn{5}{l}{\textit{Contextualization}} \\
    \addlinespace[0.25em]
    \quad Indirect via ET                & $-.004$ & .009 & [$-.022$, $\phantom{-}.013$] & .628 \\
    \quad Indirect via CT                & $-.021$ & .019 & [$-.059$, $\phantom{-}.016$] & .269 \\
    \quad Total indirect                 & $-.026$ & .023 & [$-.070$, $\phantom{-}.019$] & .263 \\
    \quad Direct                 & $-.011$ & .084 & [$-.175$, $\phantom{-}.154$] & .900 \\
    \quad Total                          & $-.036$ & .086 & [$-.205$, $\phantom{-}.132$] & .675 \\
    \bottomrule
    \end{tabular*}
    \appnote{Standardized effects. ET: emotional trust, CT: competent trust.}
\end{table}

\end{document}